\documentclass{article}

\usepackage{arxiv}

\usepackage{hyperref}
\usepackage{graphicx}
\usepackage[utf8]{inputenc}
\usepackage[T1]{fontenc}
\usepackage{lmodern}
\usepackage{hhline}
\usepackage{tabu}
\usepackage{xcolor}
\usepackage{dcolumn}
\usepackage{tabu}
\usepackage{tabularx}
\usepackage{booktabs}  
\usepackage{multirow}
\usepackage{rotating}
\usepackage{amsmath,amsfonts,amssymb}
\usepackage[super]{nth}
\usepackage{verbatim}
\usepackage{xspace}
\usepackage{ragged2e}
\usepackage{hyperref}
\usepackage[center]{subfigure}
\usepackage{booktabs}
\usepackage[normalem]{ulem}

\usepackage[toc,page]{appendix}

\newcommand{\stkout}[1]{\ifmmode\text{\sout{\ensuremath{#1}}}\else\sout{#1}\fi}

\usepackage{enumitem}
\newlist{abbrv}{itemize}{1}
\setlist[abbrv,1]{label=,labelwidth=1in,align=parleft,itemsep=0.1\baselineskip,leftmargin=!}

\newcommand{\orcid}{\textsf{ORCID}\xspace}
\newcommand{\hits}{\textsf{HITS}\xspace}
\newcommand{\smn}{scientific migration network\xspace}
\newcommand{\smnshort}{\textsf{SMN}\xspace}

\providecommand{\keyword}[1]{\textbf{\textit{Keywords:}} #1}
\usepackage[figuresfirst]{endfloat}

\begin{document}

\title{Measuring Scientific Brain Drain with Hubs and Authorities: a Dual Perspective}

\author{
Alessandra Urbinati \\
  Dipartimento di Informatica\\
  Universit\`a degli Studi di Torino\\
  Torino, Italy \\
  \texttt{alessandra.urbinati@unito.it} \\
   \And
Edoardo Galimberti \\
  Dipartimento di Informatica\\
  Universit\`a degli Studi di Torino\\
  Torino, Italy \\
  \texttt{edoardo.galimberti1@gmail.com}
  \And
Giancarlo Ruffo \\
  Dipartimento di Informatica\\
  Universit\`a degli Studi di Torino\\
  Torino, Italy \\
  \texttt{giancarlo.ruffo@unito.it} \\
}

\maketitle

\begin{abstract}
\justify
We studied international migrations of researchers, scientists, and academics, to better understand the so-called ``brain drain'' phenomenon, if and how it can be measured, and how it changes over time. We discuss why some trivial measures can be ineffective, and as a consequence, we built the global scientific migration network to identify the most important countries involved in the mobility of scholars, and to study their role at a local and a global scale. 

For such a purpose, we analysed a temporal directed weighted network representing scientists moving from one country to another, from 2000 to 2016, built on top of $2.8$ million \orcid public profiles. With the support of the well-known \hits algorithm, we found \emph{hubs} and \emph{authorities} to study the interplay between providing and attracting researchers from a global perspective, and its relationship to other structural features.

Our findings highlight the presence of a set of countries acting both as hubs and authorities, occupying a privileged position in the Scientific Migration Network, that is network of the scientific migrations, and having similar local characteristics, i.e., several neighbours with highly differentiated flows of researchers moving from/to them. However, it is striking that some of these countries have a predominant role over the others, and that we can easily observe countries that are extremely more attractive than others, as well as other countries that perform better as exporters than importers of scientists. It is also interesting that hubs and authorities scores can change over time, alongside with their relative discrepancy, and other network measures, suggesting that local and/or global policies can buck the trend. 

\end{abstract}

\keyword{scientific migration, complex networks analysis, hyperlink-induced topic search, science of science}


\section{Introduction}
Human migration has always been a phenomenon of crucial importance in history and it has radically evolved over time, affected by historical and economic events.
It is known for shaping local demographics, politics, and regulations; and, also, for influencing global wealth and world-wide society~\cite{organisation2008profile},~\cite{klugman2009human}.

The definitive outcome of human migration is subtle and extremely unpredictable, especially in the long term, due to the need for addressing different borders: geographical, political, and even cultural~\cite{rinzivillo2012discovering}.
For these reasons, human migration is perceived in many different manners and, consequently, treated by local states with opposite aims: it is sometimes encouraged, rather discouraged~\cite{schiantarelli2005global}. In particular, knowledge, ideas, and information are considered to be among the most relevant assets in today's economy and are naturally embedded in researchers, scientists, and academics who, through their permanent or temporary mobility paths, move such goods from a location to another~\cite{moed2013studying}.
On the long term, international scientific mobility could impact fundamental social and economic aspects of the countries, such as scientific, technological, and productive assets~\cite{pugliese2017unfolding}. Please, observe that hereinafter the terms ``mobility'' and ``migration'' will be used interchangeably to indicate the event of a researcher moving from one country to another, without differentiating permanent or long stays from short stays such and scholarships or post-doc periods.
Albeit, most of the times, this phenomenon lacks the urgency of survival, it is highly competitive in terms of choice of the destination countries, as pointed out in~\cite{deville2014career}.

In this paper, we want to explore scientific migration as a global and inherently interdependent phenomenon. We analyse different frameworks to detect those countries that better attract or repel researchers, to characterise different roles, and to understand how mobility dynamics change over time: the so-called ``brain drain'' phenomenon. We rely our analysis on \orcid, a growing platform that collects public profiles of researchers.In particular, we employed $2.8$ million public profiles created until 2016 and already used for other preliminary analysis~\cite{bohannon2017introducing}. 
Given its nature, the (scientific) migration system can be modelled using a network that we define to be temporal, weighted, and directed: it turns out that a complex network perspective is very useful to define relationships between actors involved in this ecosystem, and it also provides a solid ground to define measures and parameters that can be used to study efficiently the mobility phenomenon. In this domain, nodes represent world countries and edges account for a migratory flow from a country to another.
Edge weights stand for the size of the migratory flow in terms of migrants, while timestamps represent years from 2000 to 2016.
We name such structure \emph{\smn}\ (\emph{\smnshort}\ for short).

In our setting, we aim at identifying those countries that are able to provide or attract a large number of outgoing or incoming researchers.
Apparently, these characteristics are antithetical and they are worth to account separately. However, in principle, every single node in a directed network can outperform according to both their in-degree and out-degree (or also by their in-strength and out-strength in weighed networks), so we need a measure that allows for nodes to play both roles in the mobility ecosystem. Also, and most importantly, we cannot neglect the global interdependence of the migration phenomenon, and that mobility cannot be simplified in terms of the number of researchers that move from one country to another, neglecting that this can be just one step of a longer path involving many different nodes. In fact, to capture these characteristics, we employ the well-know Kleinberg's \emph{weighted hyperlink-induced topic search} (\hits) algorithm on the \smn to identify \emph{authorities} and \emph{hubs}~\cite{kleinberg1999hubs}.
We compare the results obtained by \hits to other local and global measures, to show that that a dual perspective based on hubs and authorities provides more insights to unfold the interplay between exporting and importing researchers on large scale.
Further, we investigate the local patterns and characteristics of successors of hubs and predecessors of authorities to derive the motivations behind the \hits\ algorithm.

Our results show a high correlation between hub and authority countries.
In particular, we are able to identify a set of countries that occupies a privileged position in the \smn, being both important hubs and central authorities, since they are able to receive researchers and, at the same time, to provide scientists to the most attractive countries.
This finding probably contradicts the common perception that countries attracting researchers are not good providers, and vice versa.
Also, we observe that heterogeneity in the local neighbourhood leads countries with very different social and economic background to reach similar hub or authority scores over the years. 
External factors, e.g., regulations, political alliances, investments in research, development, and education, are expected to play an important role in such results and to add an additional layer of complexity that deserves to be investigated further. 

\section{Related work}
\label{sec:background}
In this section we give an overview of existing related literature, also to better introduce the main contributions of this paper. 

\subsection{\orcid\ data}
To our knowledge, the first attempt to utilise \orcid\ data in order to extract meaningful information about the migration of the scientific population has been carried out in~\cite{bohannon2017introducing}.
The authors first claim that, despite having biases, \orcid\ data can be used to survey scientific migration given the high adoption rate by the academic population.
Then, they provide a collection of basic statistics about the dataset without deepening the temporal evolution of the phenomenon nor adopting a network approach.

\subsection{Network analysis of human migration}
Human migration is modelled in terms of complex networks is~\cite{fagiolo2013international}.
Similarly to our case, they define the \emph{international migration network} as a temporal weighted directed network having countries as nodes and volumes of migrants as edges.
Differently from our work, the study by Fagiolo~\emph{et.~al.} mostly focuses on the identification of community structures and disassortativity; 
moreover, it considers the general human migration that has fundamentally different characteristics than the scientific one.
Following up this seminal work, many other approaches are proposed with similar purposes, studying for example human migration from a multi-layer perspective using data gathered from social media platforms~\cite{belyi2017global}.
A complementary work~\cite{fagiolo2015human} correlates per-capita income and labour productivity with human migration and network centrality.
It has been explored also how to build complex networks from worldwide migration flows to identify a socioeconomic indicator that explains the reasons behind the phenomenon~\cite{cerqueti2018network}.
Robinson~\emph{et.~al.}~\cite{robinson2018machine} propose a machine learning approach to predict long-term human mobility.
Finally, other works, as~\cite{guidotti2016unveiling}, employ the network structure to unfold information about human mobility from GPS and GSM data. 

\subsection{Scientific migration}
The mobility of scientists is a topic of broad interest that has been investigated in a series of works.
The mobility of scientists within and across countries is studied in~\cite{geuna2015global} adopting an economic point of view mixed with the traditional sociology of science.
Saxenian~\cite{saxenian2005brain} and Agrawal~\emph{et.~al.}~\cite{agrawal2011brain} discuss the concept of \emph{brain drain} and argue that connections between migrant scientists and their home countries are persistent in time and might ease knowledge transfer backward.
For these reasons, they call this phenomenon \emph{brain circulation} or \emph{brain bank}.

Since reliable data sources about the topic are often problematic, a survey has been devised in~\cite{franzoni2012foreign} with the intent of providing consistent data about cross-country researches. The study documented in~\cite{moed2013studying} explores how Scopus\footnote{\url{https://www.scopus.com}} can be exploited as data source to understand international scientific mobility for countries with high adoption of the platform. In the study, the authors show quantitative metrics and general trends about the observed countries and researchers.

A recent study by Verginer~\emph{et.~al.}~\cite{verginer2018brain} describes a method to extract mobility networks from a collection of four bibliographic data sources, not including \orcid, to characterise the mobility of scientists at city granularity, finding evidence that global cities attract highly productive scientist early in their careers.

\subsection{Applications of the \hits algorithm beyond the Web}
Although \hits was initially proposed to better identify the most important Web pages related to a given topic~\cite{kleinberg1999hubs}, it has been proved to be applicable in many different domains. For instance, the authors of~\cite{deguchi2014hubs} investigated the economic hubs and authorities of the world trade network in time using the \hits\ algorithm.
On the other hand, the \hits\ algorithm is applied in~\cite{safavi2018career} to a career network for studying careers path of Ph.D.s in Computer Science and for understanding the flow of expertise and talent across organisations.
\hits\ can be extended also to a multi-layer framework, as shown in \cite{bonaccorsi2019country}, that investigated how the centrality of a country correlates to the GDP per capita.

\section{Dataset and network model}
\label{sec:dataset}
\subsection{Dataset}
The dataset employed in this work has been assembled by Bohannon and Doran~\cite{dryad_48s16} through the gathering of 2.8 millions \orcid public profiles.
\orcid is a nonprofit organisation that collects contributions, affiliations, and personal information of the subscribed researchers.
Given the affiliation history of each member, we are able to identify the location, in terms of country, of their workplace over time and infer scholars' migration across different states.
In the following paragraphs, we introduce the dataset on annual basis to ease of interpretation and due to data limitations, i.e., the temporal information inserted by the users often lacks of month granularity.

\begin{figure}[ht!]
\centering
\includegraphics[width=0.95\columnwidth]{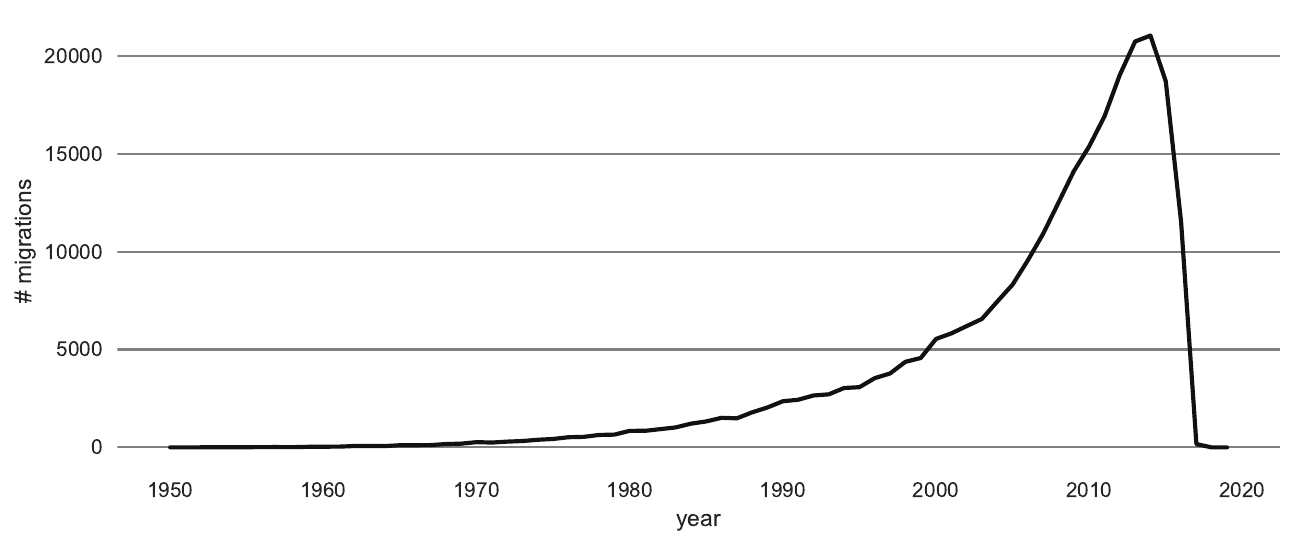}
\caption{Distribution of the number of \orcid\ members migrating per year, from 1950 to 2020.}
\label{fig:data_distr}
\end{figure}

Figure~\ref{fig:data_distr} shows the distribution of the number of estimated migrations, i.e., the number of \orcid\ members that edited their profile to change the country they worked in, per year, from 1950 to 2020.
Most of the data is concentrated in the \nth{21} century, with a peak in 2014.
The decay of recorded migrations after 2014 might be due to temporal bias given by the time when the dataset was gathered, i.e., in 2017.
Even if \orcid\ was founded in 2012, members are allowed to insert information about their previous occupations and their planned ones; as a consequence, we have data about migrations that happened before 2012 and to occur after 2017.

In their work, Bohannon and Doran~\cite{bohannon2017introducing} highlight that \orcid\ was not designed with the specific aim of tracking researchers' mobility.
Therefore, the data we consider has structural limitations as well as biases.
First of all, as already observed, much of the information created by the members is retroactive since it refers to periods preceding  \orcid\ 's launch in 2012.
Therefore, some of the countries that nowadays have changed their political-geographical characteristics, are present in the dataset, making the set of considered countries highly variable year after year.
Secondly, since its appearance, \orcid\ has always focused mainly to younger researchers.
In fact, new subscriptions are often referred to researcher that pursued their Ph.D. recently, creating an over representation of this category in the dataset, and reflecting the fact that younger researchers sign-up to \orcid\ more frequently than elder ones.
Finally, countries are not equally represented, namely, the distribution of the number of researchers per country does not follow the distribution of the overall population.
Bohannon and Doran compare \orcid\ data in 2013 about scientific migrations to the \textsf{UNESCO} Science Report\footnote{\url{https://en.unesco.org/node/252273}} to discover which countries are misrepresented; e.g., China, Russia, and Japan result to be under represented while, e.g, Spain, and Portugal are over represented.
For these reasons, we cannot regard the dataset as a definitive picture of the scientific migrations.
Nevertheless, we can exploit it to detect regularities and patterns by the construction of a network model, useful in the understanding of the global perspective of the phenomenon, suggesting that experiments and estimations should be re-executed periodically to better monitor the phenomenon, tune previously introduced errors due to misrepresentation, and update information with fresh new data inserted/modified by researchers.

\subsection{Data processing}
\begin{figure}[ht!]
\centering
\includegraphics[width=0.95\columnwidth]{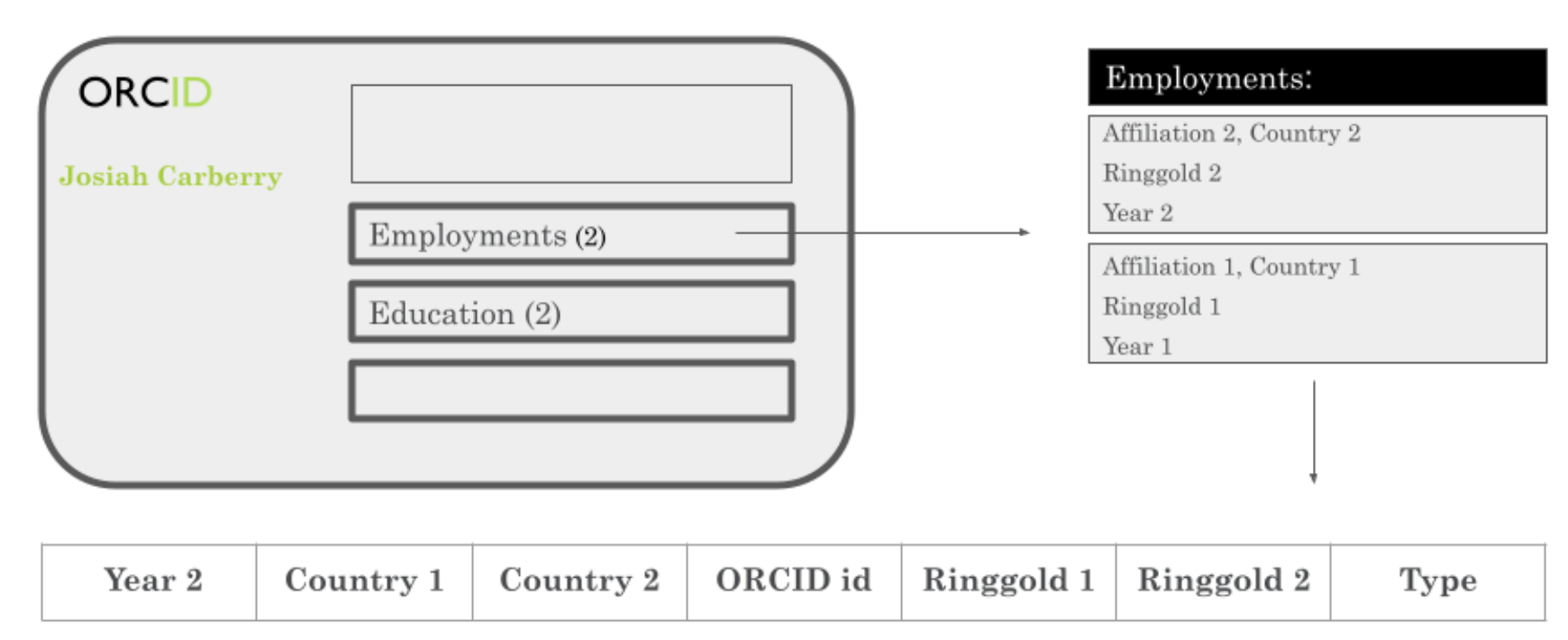}
\caption{Pipeline of data preparation: from row data to network data. Josiah Carberry is a fictitious person, his account is used as a demonstration account by ORCID.}
\label{fig:data_orcid}
\end{figure}
The raw dataset of \orcid database is a collection of files, one for each user that has decided to utilise the platform. As shown in Figure~\ref{fig:data_orcid}, we scan every user file and collect all the affiliation's changes at a yearly level, gathering both education and employment movements. A scientific migration happens if the country of one of these two affiliations changes. ``Ringgold'' labels account for the specific institute of the affiliation. The label ``Type'' retains the information about the nature of the migration. It combines two domains, ``xy'', where both can assume the values education (``ed'') or employment(``em''). The set-up of the string allows us to interpret the reason for the researcher's migration, for example from education to employment: ``edem''. In the current work, we have not employed the different reasons behind migration, but we plan to investigate the matter further in future works \footnote{For questions regarding data processing write directly to the corresponding author.}.

\subsection{Network model}
We consider a weighted directed temporal network $G = (V,T,\varpi)$,
where $V$ is a set of nodes, $T = [t_0, t_1, \ldots, t_{max}] \subseteq \mathbb{N}$ is a discrete time domain,
and $\varpi: V  \times V \times  T\rightarrow \mathbb{N}$ is a function defining for each pair of nodes $i,j \in V$ and each timestamp $t \in T$ the weight of edge $(i,j)$ at time $t$.
In the following, we refer to the weight of edge $(i,j)$ at time $t$ as $w_{ij,t}$, and we consider it missing if $w_{ij,t} = 0$.
Let $s_{i,t}^{in} = \sum_{j \in V} w_{ji,t}$ and $s_{i,t}^{out} = \sum_{j \in V} w_{ij,t}$ represent the in-strength and the out-strength of node $i \in V$ at time $t \in T$, respectively.
We also denote by $E_t = \{(i,j) \mid \varpi(i,j,t) > 0 \}$ the set of edges existing at time $t \in T$.
Finally, let $W_t$ be the weighted adjacency matrix of $G$ at time $t \in T$.

In our application domain, we identify the nodes of the network as the countries involved in the scientific migration process (231 in total); an edge between two countries represents a migration route.
Each edge between two nodes $i,j \in V$ is attributed with a time $t \in T$ and a weight $w$: a quartet $(i, j, t, w)$ represents the migration of $w$ researchers from country $i$ to country $j$ at time $t$.
The time domain of the \emph{\smn} is $T = [2000, 2001, \ldots, 2016]$, composed of 17 years, since most of the data is concentrated between 2000 and 2016, and the geopolitical configuration of the countries is quite stable after 2000. 2014 is  the year for which the dataset records the largest amount of information, so we consider it pivotal in the following analysis.

\section{From methods to measures}
\label{sec:role}
\subsection{A strength-based approach}
A strength-based approach can be considered a straightforward attempt to numerically quantify the role of a country in the \smn.

\begin{figure}[ht!]
\centering
\includegraphics[width=0.98\columnwidth]{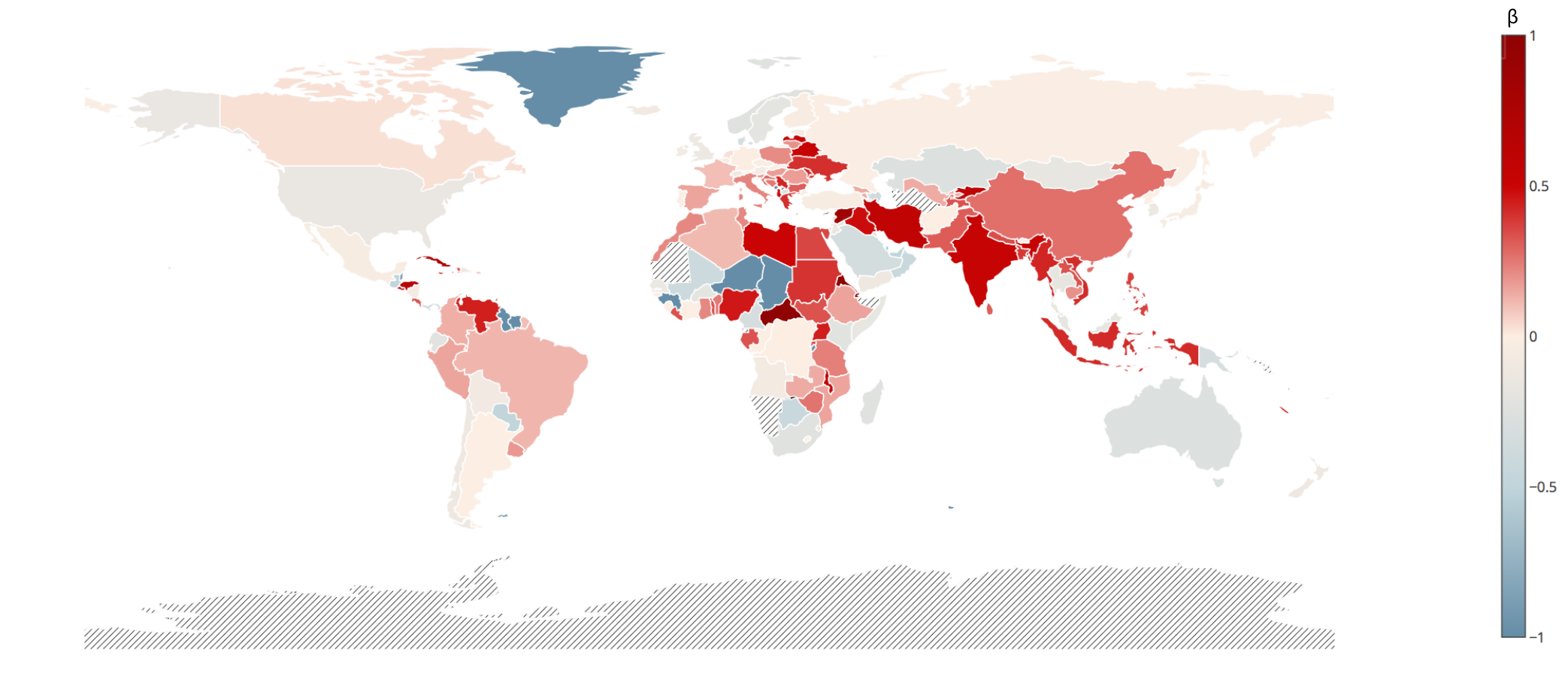}
\caption{Drain index $\beta$ in 2014. Positive (negative) values of $\beta$ are colour coded with different shades of red (blue). Countries without data have been dashed with diagonal lines.}
\label{fig:drain_index}
\end{figure}

We can intuitively define the \emph{drain index} of a country $i \in V$ at time $t \in T$ as
\begin{equation}\label{eq:brain_index}
\beta(i,t) = \frac{s_{i,t}^{out} - s_{i,t}^{in}}{s_{i,t}^{out} + s_{i,t}^{in}},
\end{equation}
namely the number of outgoing researchers (i.e., out-strength) minus the number of incoming researchers (i.e., in-strength) normalised by their sum.
It ranges from $-1$ to $1$, where $1$ indicates maximum brain drain (the country is a pure provider) while $-1$ means maximum brain gain (the country is a pure receiver).
Values close to $0$ are adopted by those countries having balanced values of out-strength and in-strength.

\begin{table}[ht!]
\centering
\caption{Ranking (partial) of the countries by drain index $\beta$ in 2014.
For each country, out-strength and in-strength measured during such year are also reported.
Countries highlighted in bold have the highest out-strength in 2014.
\label{tab:drain_index}}
\begin{tabularx}{0.95\columnwidth}{cXccc}
\toprule
ranking & country & $\beta$ & $s^{out}$ & $s^{in}$\\
\midrule
$1$ & Sint Maarten & $1.0$ & $2$ & $0$\\
$2$ & Eritrea & $1.0$ & $2$ & $0$\\
$3$ & Central African Republic & $1.0$ & $1$ & $0$\\
$4$ & Curacao & $1.0$ & $1$ & $0$\\
$5$ & Saint Vincent & $1.0$ & $1$ & $0$\\
$\mathbf{85}$ & \textbf{Spain} & $0.03$ & $80$ & $74$\\
$\mathbf{90}$ & \textbf{United Kingdom} & $0.01$ & $109$ & $105$\\
$\mathbf{111}$ & \textbf{France} & $0.0$ & $78$ & $78$\\
$\mathbf{114}$ & \textbf{United States} & $-0.008$ & $114$ & $116$\\
$\mathbf{116}$ & \textbf{Italy} & $-0.01$ & $71$ & $73$\\
$202$ & Guinea & $-1.0$ & $0$ & $2$\\
$203$ & Guyana & $-1.0$ & $0$ & $2$\\
$204$ & Belize & $-1.0$ & $0$ & $2$\\
$205$ & Niger & $-1.0$ & $0$ & $3$\\
$206$ & Chad & $-1.0$ & $0$ & $3$\\
\bottomrule
\end{tabularx}
\end{table}
Figure~\ref{fig:drain_index} graphically shows the drain index for the year 2014, while Table~\ref{tab:drain_index} reports the ranking for specific countries: the five countries of highest $\beta$, the five countries of lowest $\beta$, and the five countries of highest out-strength.
The countries standing out in Figure~\ref{fig:drain_index} are mainly located in Africa, southern Asia and in the Caribbean, while Europe and North America have milder colours.
Extreme values of $\beta$ are assigned when the number of migrations of a country is poor and completely unbalanced.
For example, Sint Maarten has only two outgoing migrations, resulting in $\beta = 1$, while Chad has three incoming migrations and no outgoing researchers, then its $\beta$ is $-1$.
On the other hand, those countries playing a central role in the migration network have usually $\beta$ close to $0$ due to the high number of both outgoing and incoming researchers.
This is the case of, e.g., the United Kingdom and the United States.

\begin{figure}[ht!]
\centering
\includegraphics[width=0.98\columnwidth]{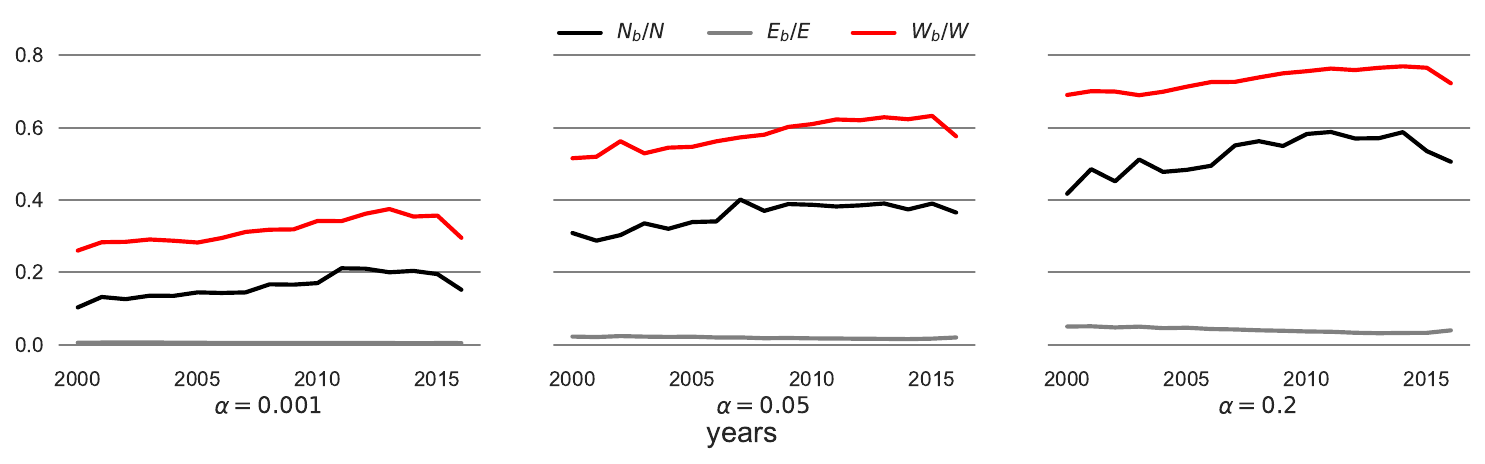}
\caption{Focus on the network backbone: figures above show the percentages of retained nodes ($N_{b}/N$), edges ($E_{b}/E$) and weights ($W_{b}/W$) after the application of the filtering strategy. Each plot shows the application of the filter with a increasing  significance levels ($\alpha$ = \{0.001, 0.05, 0.2\}).}
\label{fig:backbones}
\end{figure}

Of course, we would like to focus on countries whose the number of moving scientists is not neglectable. In order to let emerge the \emph{network backbone}, we apply the link filtering strategy that is proposed in~\cite{Serrano6483}. This operation has the aim to focus on countries that have a leading role in the scientific migration flows, while preserving the  structural characteristics of the network as a whole. Figure~\ref{fig:backbones} shows the fraction of nodes, links and weights retained by the filters according to different significance levels $\alpha$.

\begin{table}[ht!]
\centering
\caption{Countries (partial) rankings by drain index $\beta$  calculated on three different network backbones in 2014. Each backbone is extracted after the application of a filter  with a increasing  significance levels ($\alpha$ = \{0.001, 0.05, 0.2\}).
The five countries of highest $\beta$ (ties broken by out-strength) and the five countries of lowest $\beta$ (ties broken by in-strength) are reported. 
\label{tab:drain_index_threshold}}
\begin{tabularx}{1\columnwidth}{XXX}
\toprule
$alpha = 0.001$  & $alpha = 0.05$ & $alpha = 0.2$\\
\midrule
1 Iran&1 Hungary&1 Syria\\
2 Sweden&2 Cuba&2 Serbia\\
3 Greece&3 Venezuela&3 Uruguay\\
4 New Zealand&4 Uganda&4 Jamaica\\
5 Denmark&5 Zambia&5 Rwanda\\
\ldots&\ldots&\ldots\\
38 Mexico&73 Ethiopia& 117 Macao\\
39 Austria&74 Tunisia& 118 Bolivia\\
40 Chile&75 Senegal&119 Guatemala\\
41 Russia&76 Estonia&120 Brunei\\
42 South Africa&77 Luxembourg&121 Mali\\
\bottomrule
\end{tabularx}
\end{table}

\begin{figure}[h!]
\centering
\subfigure[PARAMETRI-1][$\beta_{\alpha=0.2}$  Top]{\label{fig:lv_bdTop}\includegraphics[width=0.31\textwidth,keepaspectratio]{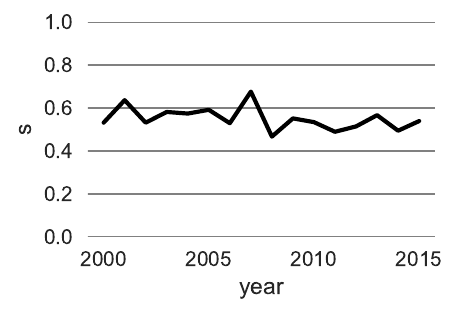}}
\subfigure[PARAMETRI-3][$\beta_{\alpha=0.2}$  Bottom]{\label{fig:lv_bdBottom}\includegraphics[width=0.31\textwidth,keepaspectratio]{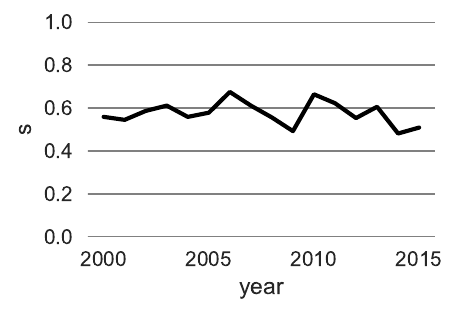}}
\subfigure[PARAMETRI-2][$\beta_{\alpha=0.2}$  Center]{\label{fig:lv_bdCenter}\includegraphics[width=0.31\textwidth,keepaspectratio]{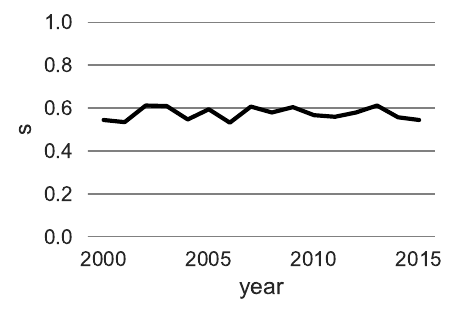}}
\subfigure[PARAMETRI-4][Page Rank]{\label{fig:lv_pg}\includegraphics[width=0.31\textwidth,keepaspectratio]{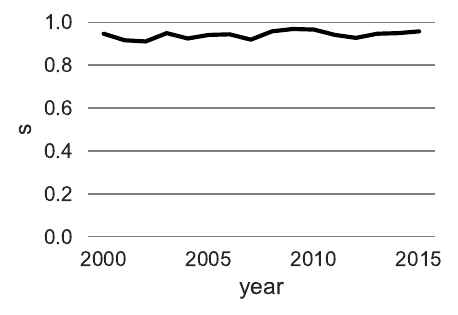}}
\subfigure[PARAMETRI-5][Authority]{\label{fig:lv_aut}\includegraphics[width=0.31\textwidth,keepaspectratio]{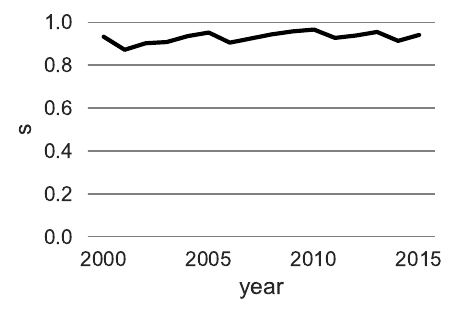}}
\subfigure[PARAMETRI-6][Hub]{\label{fig:lv_hub}\includegraphics[width=0.31\textwidth,keepaspectratio]{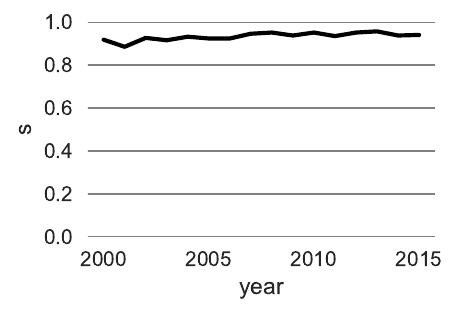}}
\caption{$s$ estimates the similarity between the rankings in two successive years. Plots in the first row represent similarities between the top twenties (a), the bottom twenties (b), and the middle twenties (c) in two successive years if we use the brain index defined in Eq.~\ref{eq:brain_index}. Plots in the bottom row represent respectively the similarities between the top 20-th countries in each ranking by page rank (d), authority score (e), and hub score (f).}
\label{fig:similarity}
\end{figure}

From the rankings displayed in Table~\ref{tab:drain_index_threshold}, and calculated on the network backbones, we intuitively observe that an high instability emerge in such rankings at varying values of $\alpha$. The ranking analysis is an open and very broad subject of interest, but a recent work~\cite{iniguez2021universal} has shown a pattern throughout its dynamics, and how for example the top part of multiple rankings shared a certain degree of stability. Also in the $\beta(i,t)$ ranking there are certain positions that carry out specific roles inside the migration system, and we would like to estimate how stable they are over the years.
To quantify it we define the Normalised Similarity $s$ between two different partial rankings $\tilde{r_{t}}$ and $\tilde{r_{t+k}}$ as:
\begin{equation}
    s(\tilde{r_{t}},\tilde{r_{t+k}})= 1 - \dfrac{1}{N(N+1)}\sum_{i \in \tilde{V_{t}}} |r_{t}(i) - r_{t+k}(i)|
\end{equation}
where $t \in T$, $k \in [1, T-t]$, and V is the set of countries that takes part to the migration network at time $t$ and occupy the chosen portion of the ranking. If at time $t+K$ a country is not the partial ranking anymore we place it in the last position of the partial ranking. The term $\dfrac{1}{N(t)(N(t)+1)}$ is an upper bound for the sum of all the possible fluctuations, in particular it would happens when all the countries at time t would downgrade at position $N+1$ while all new countries occupy the N position at time $t+k$, and  $\dfrac{(N+1)(N+1-1)}{2}+\dfrac{(N+1)(N+1-1)}{2}=N(N+1)$. In Figures~\ref{fig:similarity}(a-c) the Normalised Similarity has been computed for the key positions of subsequent rankings based on  $\beta$, calculated on the network backbone with level $\alpha= 0.2$, from the year 2000 to the year 2016. As key positions, we consider the top twenties (Fig.~\ref{fig:lv_bdTop}), the bottom twenties (Fig.~\ref{fig:lv_bdBottom}), and the twenties in the middle (Fig.~\ref{fig:lv_bdCenter}) of each ranking, that should represent respectively the top providers, the top receivers, and the most 'balanced' countries. We can easily observe that, even with a fixed value of $\alpha = 0.2$, rankings differ significantly from one year to another; in fact, $s(r_{i},r_{i+1})$ fluctuates around 0.6, meaning that the ranking calculated at year $i$ changes dramatically the following year. The lack of stability over a not-so-fast phenomenon may prevent us to spot any significant patterns or dynamics.

Additionally, we evaluated other strategies for normalising the drain index by considering external data, such as the size of the overall population and the number of researchers in a country. Given the biases in the collected dataset, any normalisation deriving from external sources would be inappropriate because it would misrepresent the results. Moreover, external data have to be temporal, at least of yearly granularity from 2000 to 2016, and available for all the countries included in the dataset. This is the case of the general population, but we cannot discover complete and coherent datasets about the size of the research population of all the studied states. However, we think that Eq.~\ref{eq:brain_index} fails mainly because it does not properly represent the complexity of the phenomenon itself: the brain index focuses on spotting 'pure receivers' and 'pure providers' in the network, whereas each country may behave accordingly a mixed streams made of scientists moving in and out. As a consequence, such a measure would suffer of a myopic view of the migration ecosystem, because it is a function of local properties only: we miss the opportunity to assess which is the role of a global and heterogeneous structure of the migration network. This is the reason why we propose the application of eigenvector centrality based algorithms to produce rankings more adequate to comparisons~\cite{alonmoshe05}.

\subsection{A global approach}
A classic approach to assess the importance of a node in a network taking into account the global link structure is the well-known \emph{PageRank}~\cite{page1999pagerank}.
\begin{table}
\centering
\caption{Top-20 ranking by PageRank in 2000, 2014, and 2016.
\label{tab:pr}}
\begin{tabularx}{0.95\columnwidth}{cXXX}
\toprule
ranking & 2000 & 2014 & 2016\\
 &  & \scalebox{0.9}{$s(r_{2000},r_{2014})=0.88$} & \scalebox{0.9}{$s(r_{2000},r_{2014})=0.91$}\\
\midrule
$1$ & United States & United States & United States\\
$2$ & United Kingdom & United Kingdom & United Kingdom\\
$3$ & Germany & Australia & Australia\\
$4$ & Spain & Spain & Germany\\
$5$ & Italy & Germany & Spain\\
$6$ & France & China & China\\
$7$ & Canada & France & Canada\\
$8$ & Australia & Canada & France\\
$9$ & Portugal & Italy & Switzerland\\
$10$ & Netherlands & Sweden & Sweden\\
$11$ & Sweden & Portugal & Netherlands\\
$12$ & Japan & Brazil & Italy\\
$13$ & Switzerland & Switzerland & Denmark\\
$14$ & Brazil & Netherlands & Portugal\\
$15$ & China & Denmark & Japan\\
$16$ & South Korea & India & Ireland\\
$17$ & Malaysia & Japan & Colombia\\
$18$ & Mexico & South Korea & India\\
$19$ & Denmark & Belgium & Brazil\\
$20$ & Indonesia & Saudi Arabia & New Zealand\\
\bottomrule
\end{tabularx}
\end{table}
Let $R_t$ be the PageRank matrix of $G = (V,T,\varpi)$ at time $t \in T$, defined as
\begin{equation}
r_{ij,t} = d \frac{w_{ij,t}}{\sum_{j \in V} w_{ij,t}} + (1-d) \frac{1}{|V|},
\end{equation}
where $d = 0.85$ is the dumpling factor.
Note that, in this work, we consider the edge weights in the definition of $R_t$.
The PageRank vector $\vec{r}_t = (r_{1,t},\ldots,r_{|V|,t})^\intercal$ is obtained by repeating the iteration
\begin{equation}\label{eq:pr}
\vec{r}_t(x+1) = R_t^\intercal\vec{r}_t(x)
\end{equation}
until convergence, with initial conditions $r_{i,t}(0) = \frac{1}{|V|}$.
$\vec{r}_t$ is computed for each timestamp, i.e., year, $t \in T$.
In the following, we often refer to the PageRank vector as $\vec{r}$ neglecting the subscript.

\begin{figure}[ht!]
\centering
\includegraphics[width=0.98\columnwidth]{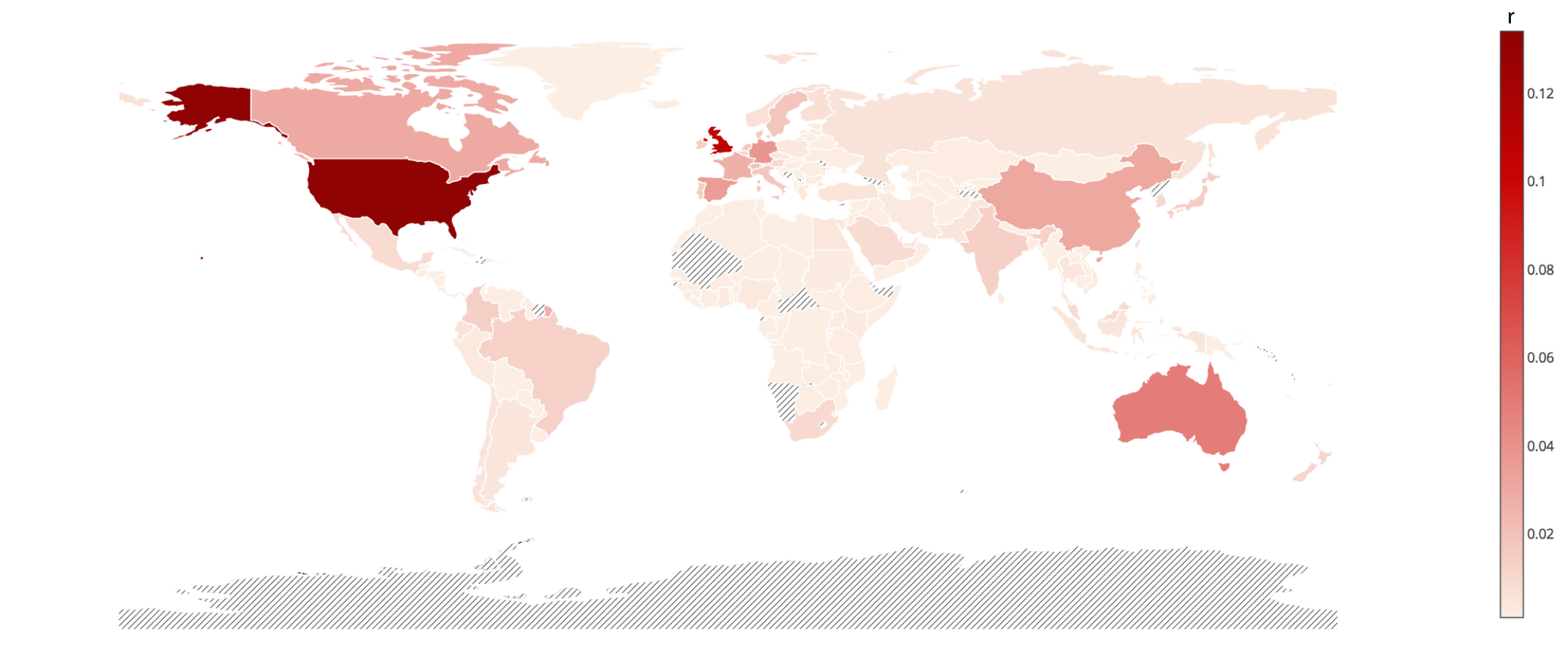}
\caption{PageRank $\vec{r}_{2014}$ is colour coded with different shades of red. Darker (lighter) red is used for countries with higher (lower) page rank values. Countries without data have been dashed with diagonal lines.}
\label{fig:pagerank_index}
\end{figure}

In Figure~\ref{fig:pagerank_index} we graphically show the PageRank in 2014, while Table~\ref{tab:pr} reports the rank of the 20 countries having highest PageRank in 2000, 2014, and 2016.
As stated above, the drain index does not privilege nodes having high both in-strength and out-strength, and does not account for the importance of the origin/destination of the connections.
PageRank is instead able to picture such aspects; in particular, United States and United Kingdom place at the first and at the second position of the ranking, respectively.

On the whole, PageRank is confirmed to be a powerful method to rank the nodes of a network, more stable according to the similarity measure $s$, as shown in Figure~\ref{fig:lv_pg} and in Table~\ref{tab:pr}.
However, it assigns to each node a unique score that is not desirable in our setting, since we are instead interested in understanding the interplay between attraction and provision of researchers.
Therefore, our analysis is required to rely on more refined and specific metrics that highlight such duality.

\subsection{A dual approach: hubs and authorities}
We identify the \emph{hyperlink-induced topic search} algorithm (also known as \hits\ or \emph{hubs and authorities})~\cite{kleinberg1999hubs} as the main measure to study our network.
The \hits\ hub vector $\vec{h}_t = (h_{1,t},\ldots,h_{|V|,t})^\intercal$ and the \hits\ authority vector $\vec{a}_t = (a_{1,t},\ldots,a_{|V|,t})^\intercal$ in $t \in T$ of $G = (V,T,\varpi)$ are defined by the limit of the following set of iterations:
\begin{equation}\label{eq:hitsh}
\vec{h}_t(x+1) = c_t(x)W_t\vec{a}_t(x+1)
\end{equation}
and
\begin{equation}\label{eq:hitsa}
\vec{a}_t(x+1) = d_t(x)W_t^\intercal\vec{h}_t(x),
\end{equation}
where $c_t(x)$ and $d_t(x)$ are normalisation factors to make the sums of all elements become unity, i.e., $\sum_{i=1}^{|V|} h_{i,t}(x+1) = 1$ and $\sum_{i=1}^{|V|} a_{i,t}(x+1) = 1$.
The initial \hits\ values of the scores are $h_{i,t}(0) = \frac{1}{|V|}$ and $a_{i,t}(0) = \frac{1}{|V|}$ for all $i \in V$.

Note that, in this work, we employ the weighted version of \hits.
The non-weighted \hits\ hub scores and non-weighted \hits\ authority scores are defined in the exactly the same way, replacing $W_t$ with the unweighted adjacency matrix in Equations~\ref{eq:hitsh}~and~\ref{eq:hitsa}.
Also in this case, $\vec{h}_t$ and $\vec{a}_t$ are computed for each timestamp, i.e., year, $t \in T$.
In the following, we often refer to the \hits\ hub and authority vectors as $\vec{h}$ and $\vec{a}$ neglecting the subscript.

\begin{table}
\centering
\caption{Best providers of scientist: top-20 ranking by hub score in 2000, 2014, and 2016.
\label{tab:hits_hub}}
\begin{tabularx}{0.95\columnwidth}{cXXX}
\toprule
ranking & 2000 & 2014 & 2016\\
&  & \scalebox{0.9}{$s(r_{2000},r_{2014})=0.87$} & \scalebox{0.9}{$s(r_{2000},r_{2014})=0.91$}\\
\midrule
$1$ & China & China & United States\\
$2$ & United Kingdom & United Kingdom & China\\
$3$ & Canada & United States & United Kingdom\\
$4$ & United States & India & Germany\\
$5$ & South Korea & Spain & India\\
$6$ & France & Canada & Spain\\
$7$ & Germany & Italy & Canada\\
$8$ & India & Germany & Italy\\
$9$ & Italy & France & Australia\\
$10$ & Spain & Brazil & France\\
$11$ & Australia & Australia & Netherlands\\
$12$ & Japan & Portugal & Brazil\\
$13$ & Brazil & South Korea & Switzerland\\
$14$ & Russia & Netherlands & Portugal\\
$15$ & Portugal & Japan & South Korea\\
$16$ & Mexico & Switzerland & Sweden\\
$17$ & Turkey & Sweden & Japan\\
$18$ & Switzerland & Iran & Denmark\\
$19$ & Colombia & Turkey & Ireland\\
$20$ & Taiwan & Colombia & Belgium\\
\bottomrule
\end{tabularx}
\end{table}

By definitions, a node $i \in V$ has large value of $h_i$ if it has many largely weighted links towards successor nodes $j \in V$ with high $a_j$;
similarly, node $i$ has large value of $a_i$ if it is reached by predecessor nodes $j \in V$ with high $h_j$ throughout largely weighted links.
In our specific scenario, $\vec{h}$ provides an indication of which are the countries playing the role of \emph{providers}, that export many researchers in direction of the most attractive countries;
while $\vec{a}$ indicates which are the \emph{attractors}, whose institutions hire researchers from highly ranked providers.

\begin{table}
\centering
\caption{Best attractors of scientist: top-20 ranking by authority score in 2000, 2014, and 2016.
\label{tab:hits_authority}}
\begin{tabularx}{0.95\columnwidth}{cXXX}
\toprule
ranking & 2000 & 2014 & 2016\\
 &  & \scalebox{0.9}{$s(r_{2000},r_{2014})=0.86$} & \scalebox{0.9}{$s(r_{2000},r_{2014})=0.88$}\\
\midrule
$1$ & United States & United States & United States\\
$2$ & United Kingdom & United Kingdom & United Kingdom\\
$3$ & Germany & Australia & Australia\\
$4$ & Italy & Germany & Germany\\
$5$ & Spain & France & Canada\\
$6$ & Canada & Canada & Spain\\
$7$ & Australia & Spain & China\\
$8$ & Portugal & China & France\\
$9$ & France & Italy & Switzerland\\
$10$ & Japan & Portugal & Netherlands\\
$11$ & Netherlands & Sweden & Sweden\\
$12$ & South Korea & Switzerland & Japan\\
$13$ & Sweden & South Korea & Italy\\
$14$ & Brazil & Netherlands & Denmark\\
$15$ & Malaysia & Brazil & Portugal\\
$16$ & Switzerland & Denmark & Hong Kong\\
$17$ & China & Japan & Ireland\\
$18$ & Ireland & Hong Kong & Colombia\\
$19$ & Mexico & India & Singapore\\
$20$ & Taiwan & Singapore & India\\
\bottomrule
\end{tabularx}
\end{table}

Tables~\ref{tab:hits_hub}~and~\ref{tab:hits_authority} show the first twenty countries ordered by hub score and authority score, respectively, in 2000, 2014, and 2016, and the similarity score $s$ between those years, whose consistency allows us some further analysis.

\subsection{Null Model}
In the rest of our analysis, we employ the \emph{configuration model}~\cite{newman2003structure} as a null model to test whether the correlation is a non-trivial feature of the \smn or if it is expected by the strength distribution of the nodes.
The configuration model rewires the edges preserving the strength distribution of the nodes in each year, namely, an edge can be shuffled only with other edges with the same timestamp. Note that by this hypothesis, in the resulting null model, the edge weight distribution and the number of edges in each year might vary with respect to the original network.
In the following results, we consider ten different configurations of the null model.

\section{Discussion}
\label{sec:disc}
To provide a more in-depth understanding of the scientific migration patterns all over the world, we focus on which are the major players that rule it, how their positions have changed over time in the ranking and inside the network structure, with the aim to detect important insight on which are the drivers that control the migration flows.
\subsection{Relationships between hub and authorities scores}
\begin{figure}[ht!]
\centering
\includegraphics[width=0.95\columnwidth]{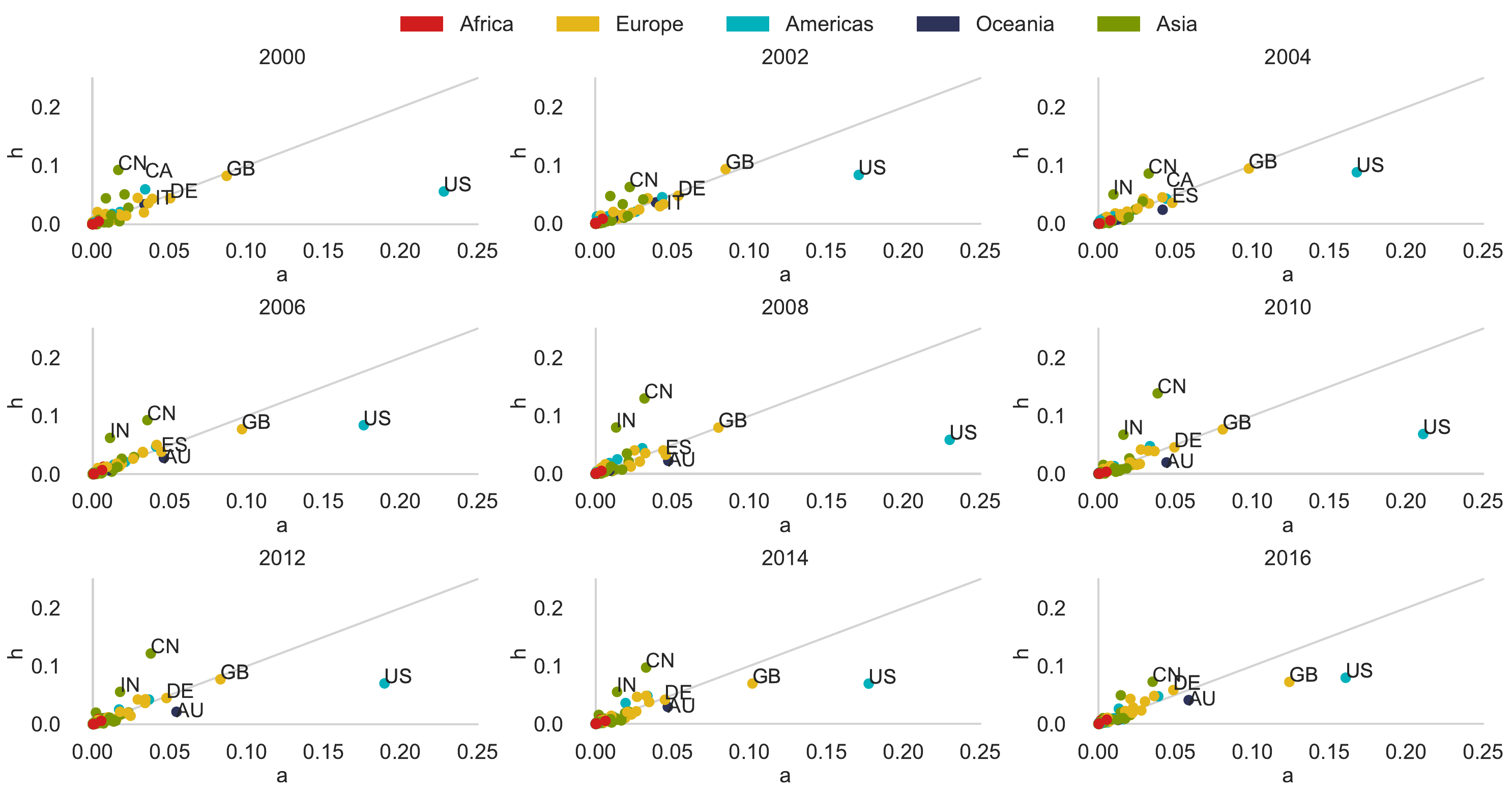}
\caption{Evolution of hub and authority scores of the nodes of the \smn in time.
ISO 3166-1 alpha-2 codes are reported for selected countries: Australia (AU), China (CN), Germany (GE), India (IN), Italy (IT), Spain (ES), United Kingdom (GB), and United States (US).}
\label{fig:hits}
\end{figure}

Figure~\ref{fig:hits} depicts the evolution of hub and authority scores of the nodes of the \smn in time, by means of scatter plots.
In all the years, most of the countries clump in the lower-left corner, where both scores are close to 0.
Most of the countries have comparable hubs and authority scores, meaning that if a country has a given role in the network as a scientists' provider, then it is likely that it has a similar role as scientists' receiver; in fact, as expected, the Pearson correlation between the two hubs/authority variables is quite high, with error always $< 1.5e\!-\!05$. However, when we calculate our scores in the null model, we find that we should have expected a higher correlation between the two variables. As a consequence, we have some outliers that buck the tends that could not have been expected with the null hypothesis, and that therefore are useful to characterise this peculiar ecosystem. Details on these comparisons are provided in~\ref{sec:B}.

Focusing on these outliers, we have that United States perform significantly better as authority than as hub, even if the corresponding hub values are always among the highest. 
On the other hand, United Kingdom moves from being equally hub and authority in early '00 to being more authority by the end of the observed period.
It is also easy to notice how China, which is constantly among the top hubs, slowly increases its authority score, with a tendency to the balance between the scores that is graphically represented by the diagonal. Such dynamics are particularly interesting, and they deserve further analysis. 

\label{sec:local}
\begin{figure}[ht!]
\centering
\subfigure[PARAMETRI-1][US Successors 2016]{\label{fig:us_succ_2016}\includegraphics[width=0.48\textwidth,keepaspectratio]{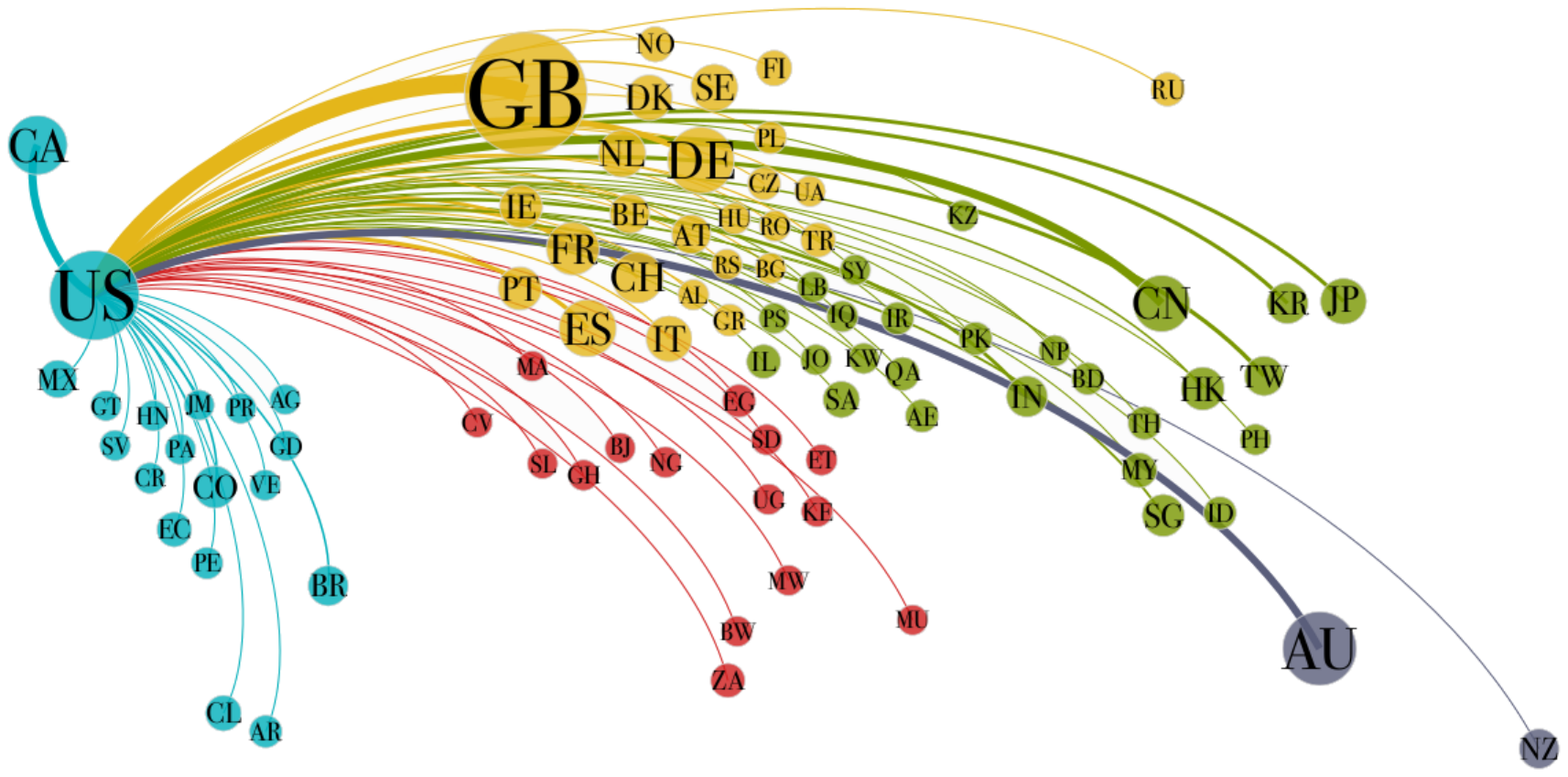}}
\subfigure[PARAMETRI-2][CN Successors 2016]{\label{fig:cn_succ_2016}\includegraphics[width=0.48\textwidth,keepaspectratio]{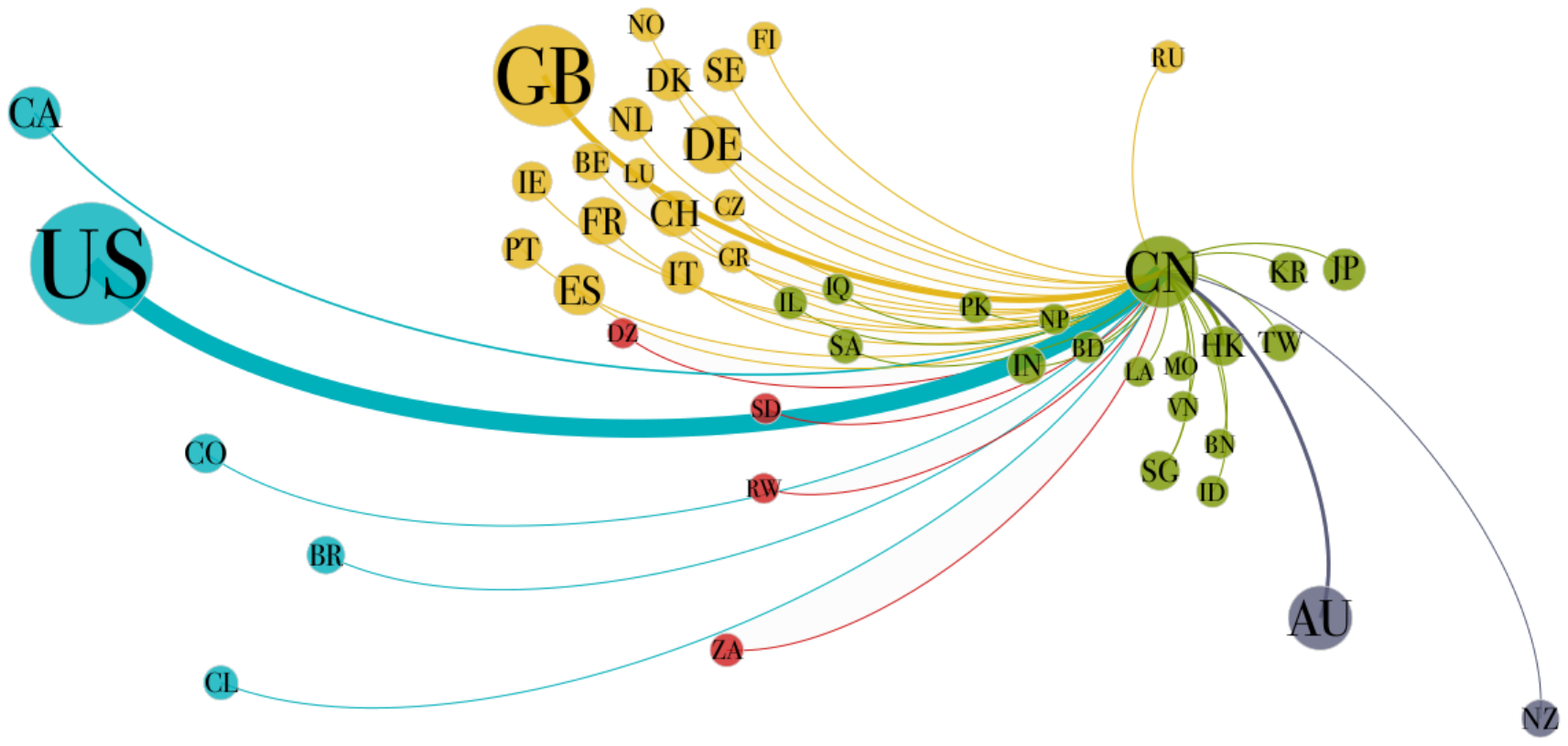}}
\subfigure[PARAMETRI-3][US Predecessors 2016]{\label{fig:us_pred_2016}\includegraphics[width=0.48\textwidth,keepaspectratio]{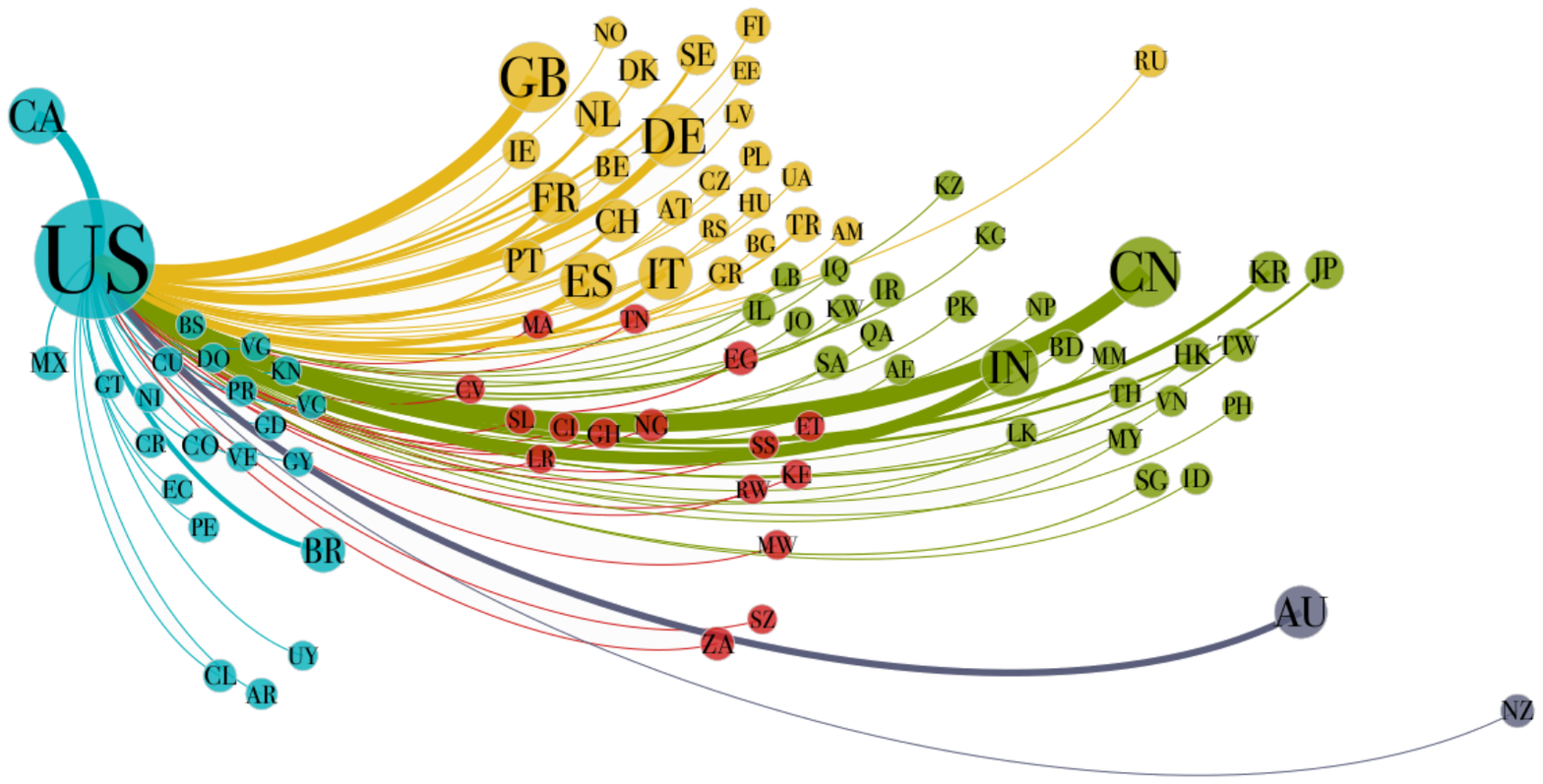}}
\subfigure[PARAMETRI-4][CN Predecessors 2016]{\label{fig:cn_pred_2016}\includegraphics[width=0.48\textwidth,keepaspectratio]{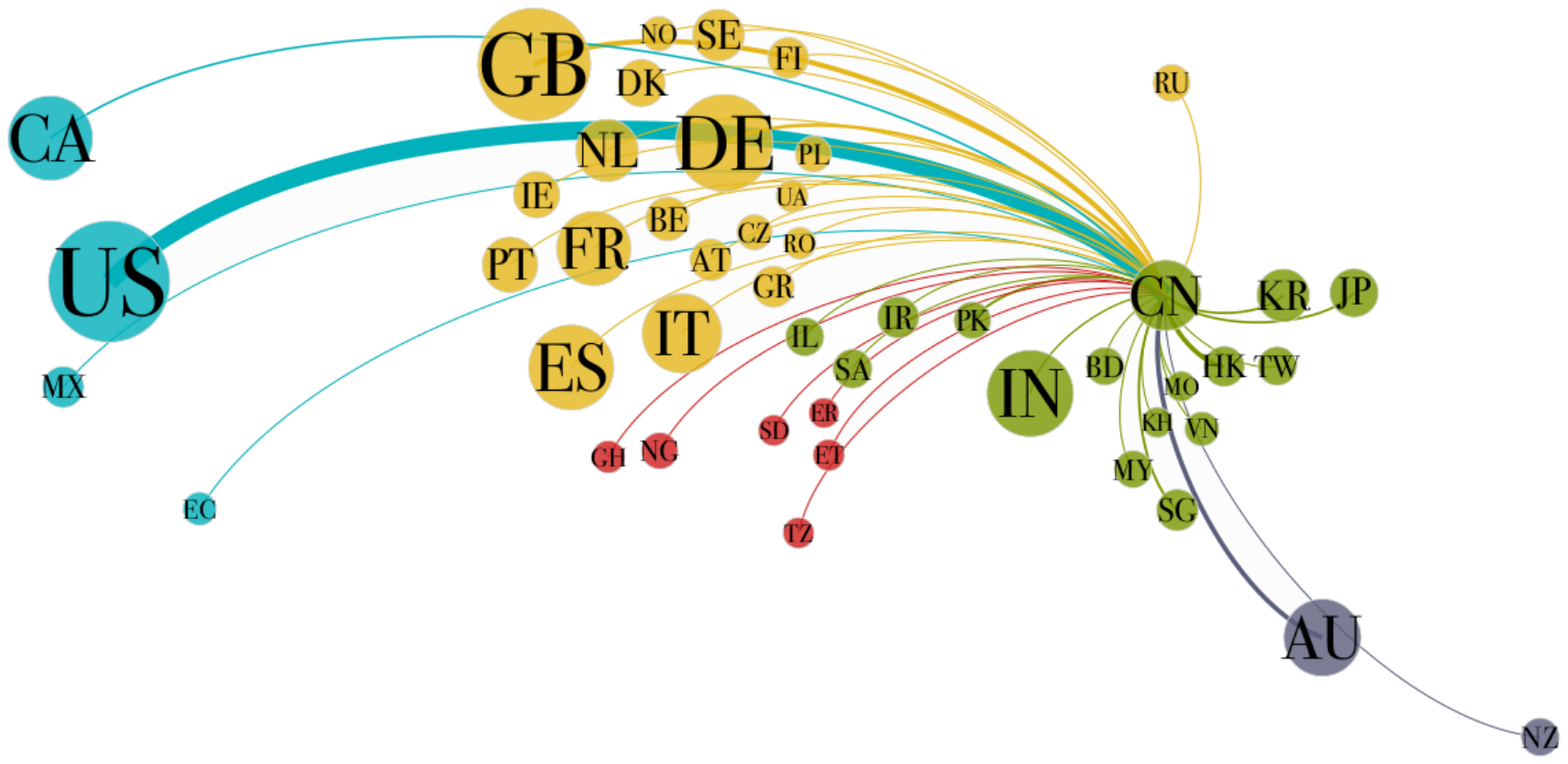}}
\caption{Evolution of the Ego-network for United States and China in 2016. Edges follow clockwise directions: we show outgoing connections (top), and incoming connections (bottom). Node dimensions scale over authority values for the attractors countries and over hub values for providers countries. Edge thickness is proportional to edge weights. Colours follow continent schema as in Figure~\ref{fig:hits}.}
\label{fig:us_cn}
\end{figure}

Figure~\ref{fig:us_cn} shows the ego-networks of the United States and China in 2016: on top there are the outgoing connections while on the bottom the incoming ones.
Colours and size of the nodes, both normalised according to each ego-network, refer to hub scores for the providers countries and authority scores for the receiving ones. Looking at the figures we notice that the United States and China both have many neighbouring countries spread across all the continents, with the United States predecessors and successors being even more scattered.  However, we are not able yet to formalise either pattern.
\subsection{Analysing local patterns with predecessors and successors}
To dive deeper into the factors that contribute to establish a country as leading hub or authority in the \smn, we investigate the homogeneity of the edge weights of the neighbourhood of the nodes.
Specifically, we want to understand how the researchers leaving (reaching) a country with high hub (authority) score is distributed over the outgoing (incoming) routes. 
In order to do so, we employ the \emph{Gini coefficient}, which measures the degree of inequality of a distribution~\cite{gini1912variabilita}.
Given a population $\mathbf{W} = \{w_o, w_1, \ldots, w_n\}$ of $n$ values, we define the Gini coefficient as
\begin{equation}
G = \frac{\sum_{w_i,w_j \in \mathbf{W}} |w_{i}-w_{j}|}{2n\sum_{w_i \in \mathbf{W}}{w_{i}}}.
\end{equation}
$G$ varies between $0$ and $1$, where $1$ expresses maximal inequality among values while $0$ indicates the case in which all the values in $\mathbf{W}$ are equal.

\begin{figure}
\centering
\includegraphics[width=0.95\columnwidth]{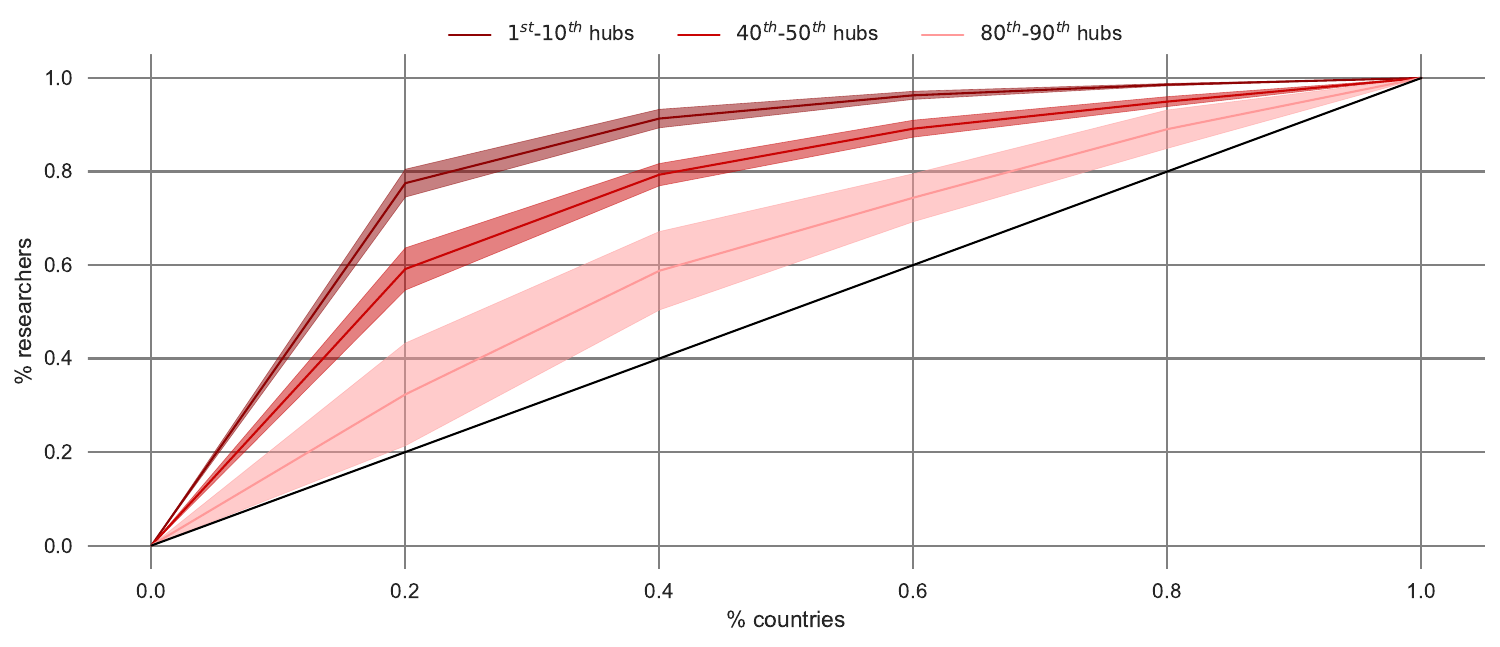}
\caption{Lorenz curves and 95\% confidence intervals for three classes of hubs in 2014.
The population $\mathbf{W}$ is represented by the edge weights of incoming edges.\label{gin_suc}}
\end{figure}

\begin{figure}
\centering
\includegraphics[width=0.95\columnwidth]{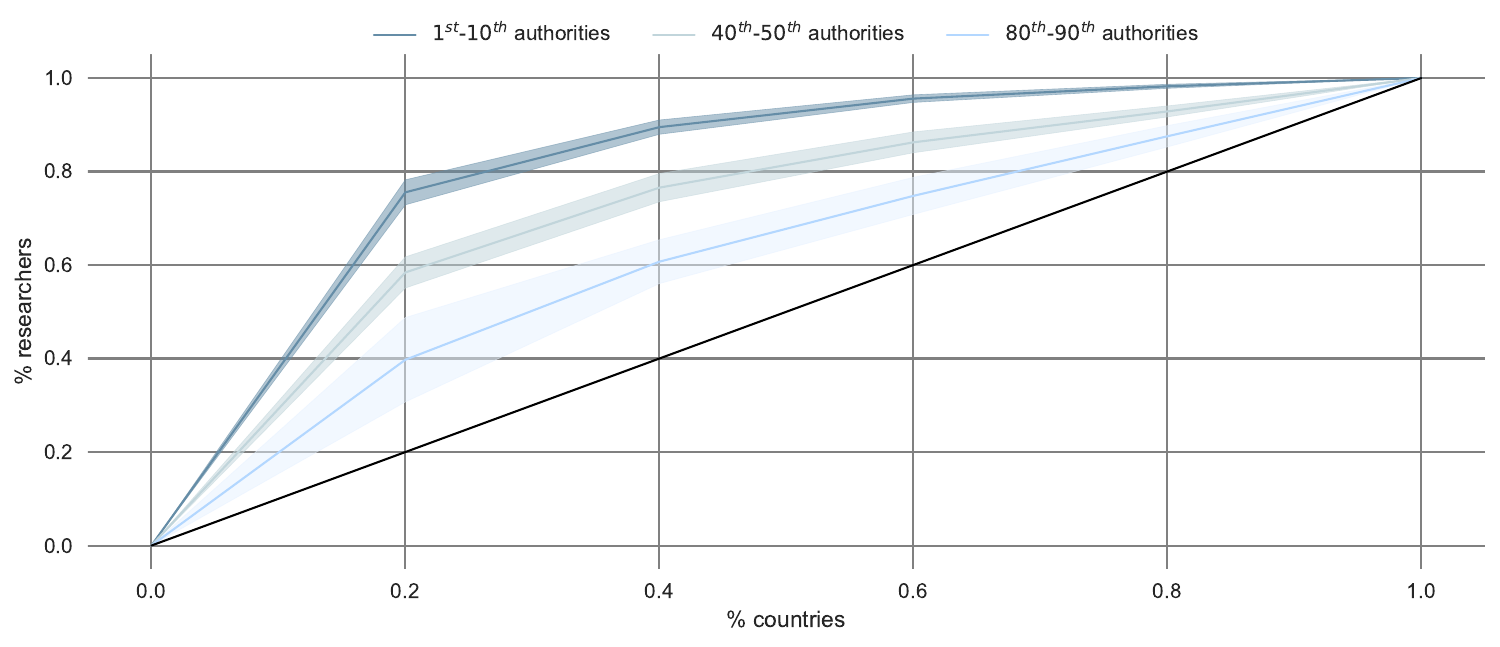}
\caption{Lorenz curves and 95\% confidence intervals for three classes of authorities in 2014.
The population $\mathbf{W}$ is represented by the edge weights of outgoing edges.\label{gin_pred}}
\end{figure}

By means of Lorenz curves it is possible to identify the population $\mathbf{W}$ as the edge weights of outgoing edges or the edge weights of incoming edges when considering a node as hub or authority, respectively.
Therefore, we aim at investigating how (un)balanced the migration flows from/towards a country are and how such aspect correlates to $\vec{h}$ and $\vec{a}$.
Figures~\ref{gin_suc}~and~\ref{gin_pred} compare the mean Lorenz curves, along with 95\% confidence intervals, of three different classes of hubs and authorities, respectively.
It is immediate to notice that high hub/authority score is associated with high Gini coefficient.
The Gini coefficient decreases progressively as we move down with the hub and authority rankings.
Then, to obtain an important position in the \smn, a country is required to have strongly differentiated migratory flows from/towards its neighbours.

\begin{figure}
\centering
\includegraphics[width=0.95\columnwidth]{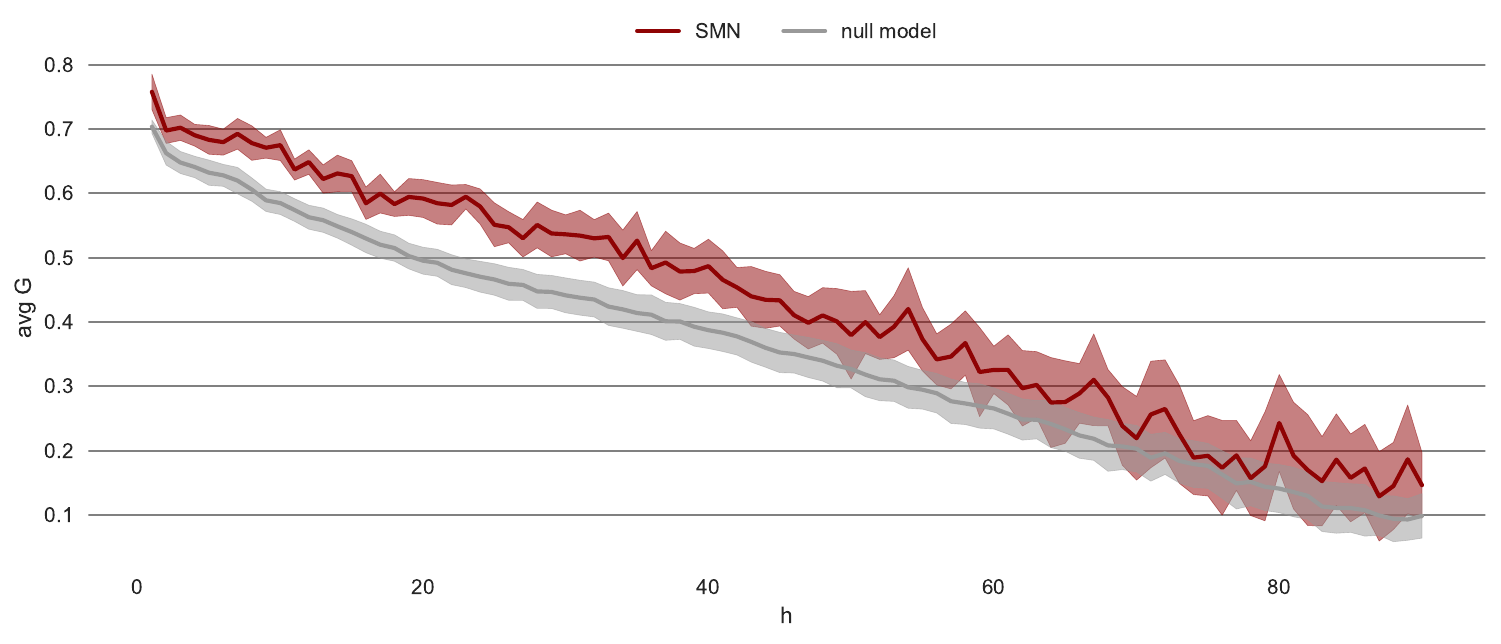}
\caption{Average Gini coefficient (and 95\% confidence interval) as a function of the hub ranking of the \smn and of the null model.
The population $\mathbf{W}$ is represented by the edge weights of outgoing edges and the average is computed over the time domain $T$.}
\label{gin_succ_cm}
\end{figure}

\begin{figure}
\centering
\includegraphics[width=0.95\columnwidth]{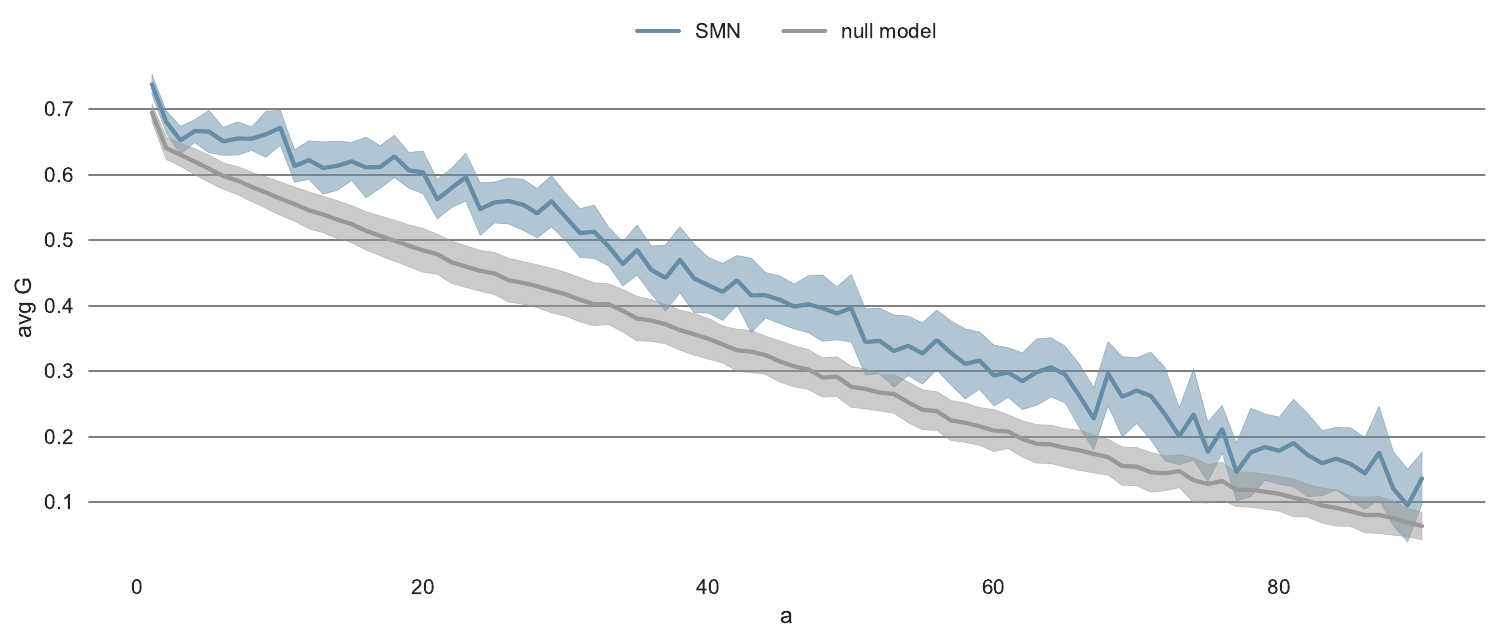}
\caption{Average Gini coefficient (and 95\% confidence interval) as a function of the authority ranking of the \smn and of the null model.
The population $\mathbf{W}$ is represented by the edge weights of outgoing edges and the average is computed over the time domain $T$.}
\label{gin_pred_cm}
\end{figure}

The behaviour of the missing classes is consistent as shown in Figures~\ref{gin_succ_cm}~and~\ref{gin_pred_cm} which report the average (over the time domain $T$) of the Gini coefficient (and the 95\% confidence interval) as a function of the hub/authority ranking.
Such curves are compared with the null model considering the average of the ten different configurations we generated.
The Gini coefficient decreases as $h$ and $a$ drop, both in the \smn and in the null model, and the curves have very similar functional shapes.
The confidence intervals are quite limited in all cases, however they become larger for the lowest positions of the ranking in the \smn where data become more sparse and less significant.
The Gini coefficient of the \smn is slightly but significantly higher than the null model; this means that a node occupying the first positions in the hub/authority ranking also shows high disparity in the weights of the connections from/to its predecessors/successors by the intrinsic characteristics of the network.

\begin{figure}[h!]
\centering
\subfigure[PARAMETRI-1][Authority Ranking]{\label{fig:aut_flow}\includegraphics[width=1.1\textwidth,keepaspectratio]{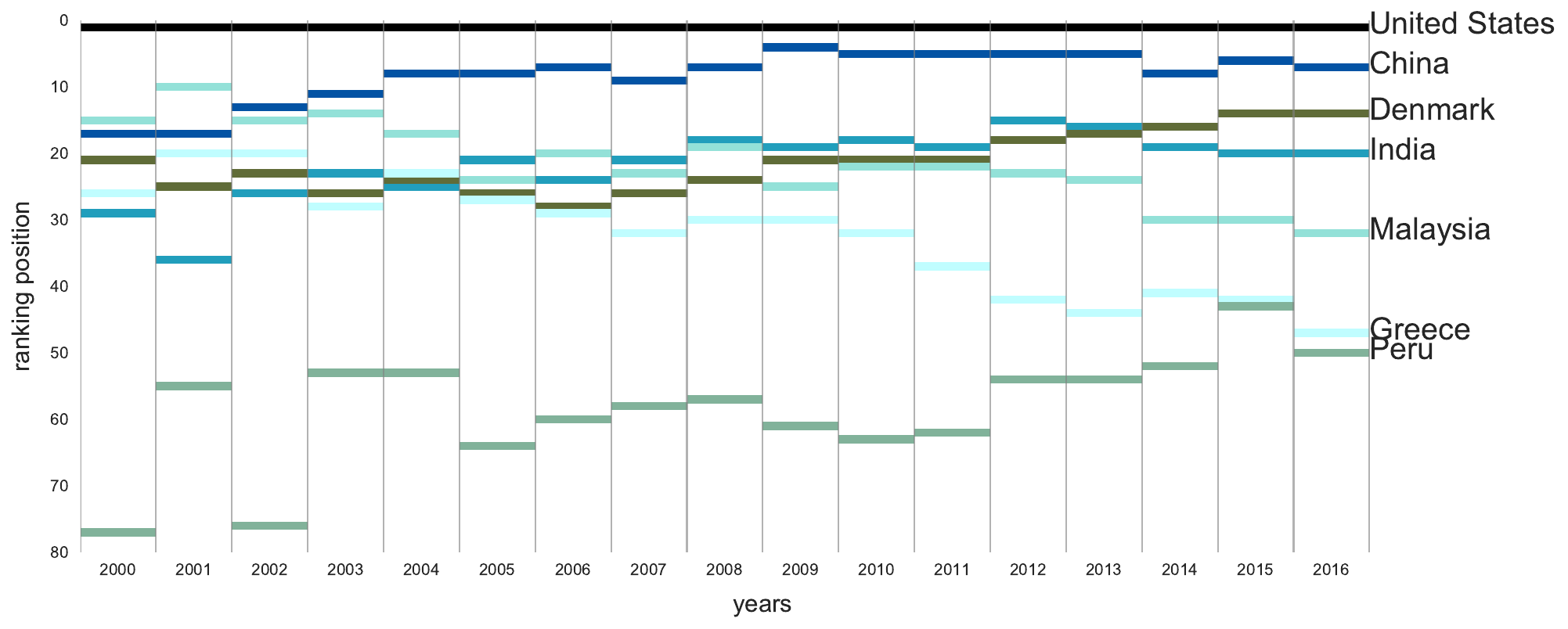}}
\subfigure[PARAMETRI-1][Hub Ranking]{\label{fig:hub_flow}\includegraphics[width=1.1\textwidth, keepaspectratio]{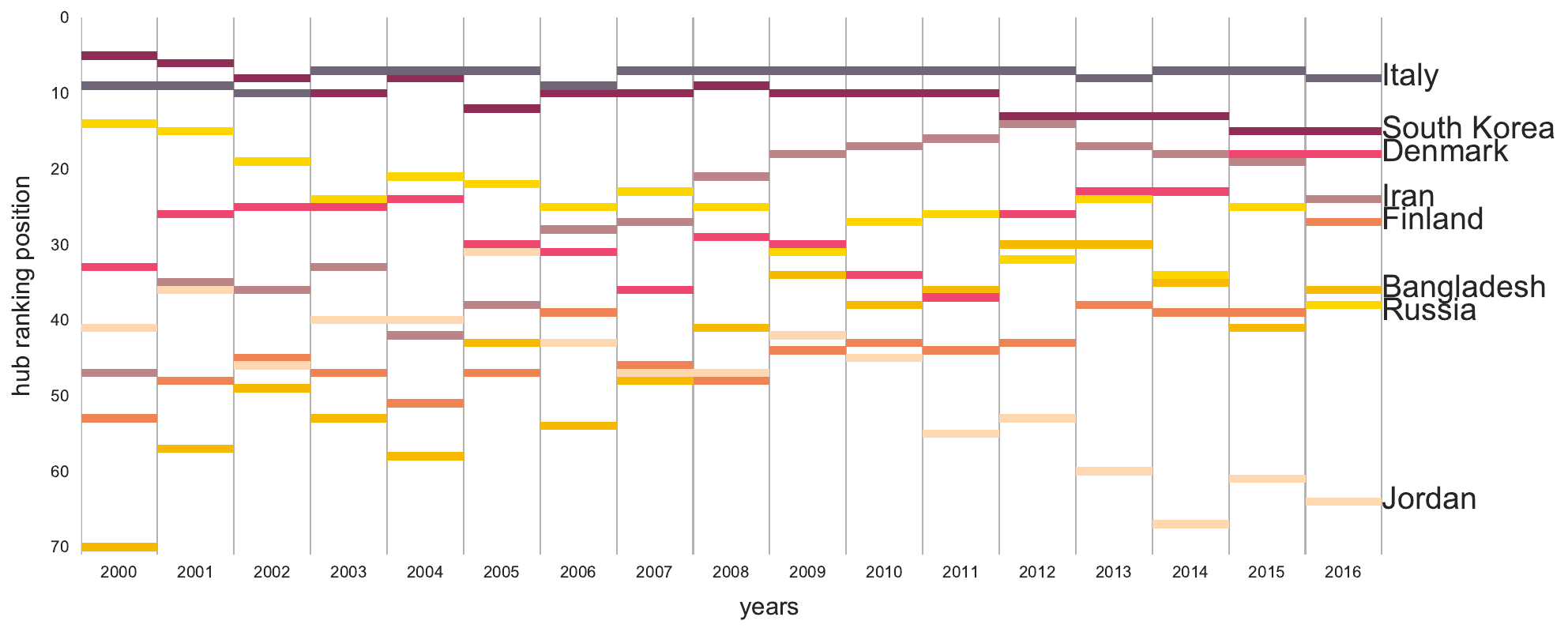}}
\caption{Ranking according to increase or decrease of position in time span 2000-2016 for authorities (a) and hubs (b).}
\label{fig:ranking_flow}
\end{figure}

\begin{figure}[h!]
\centering
\subfigure[PARAMETRI-1][Gini of weights distribution for edges to GR and PE from their predecessors]{\label{fig:gini_gr_pe}\includegraphics[width=0.45\textwidth,keepaspectratio]{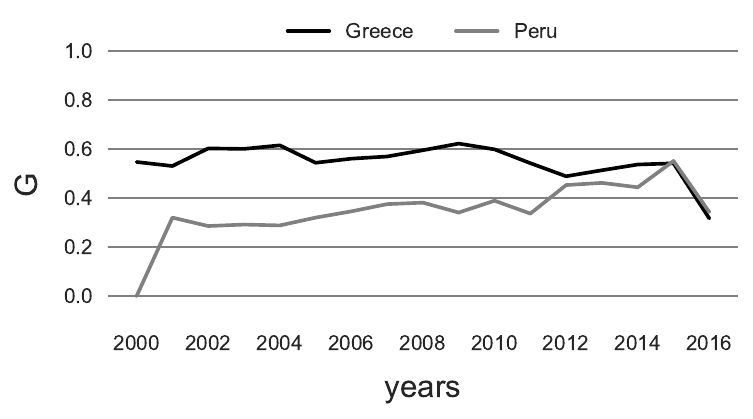}}
\subfigure[PARAMETRI-1][Gini of weights distribution for edges from SK and DK to their successors]{\label{fig:gini_sk_dk}\includegraphics[width=0.45\textwidth, keepaspectratio]{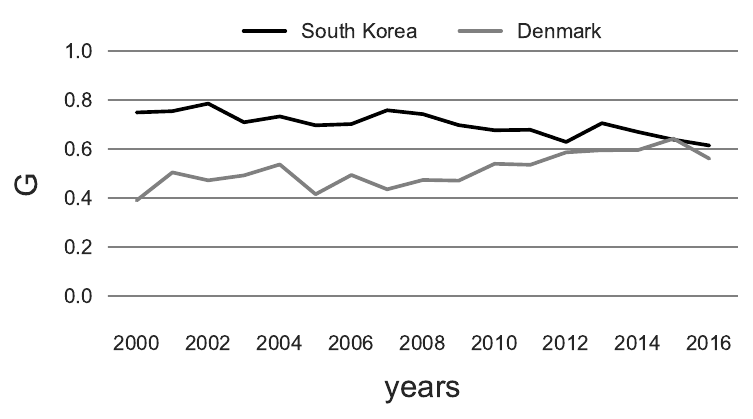}}
\subfigure[PARAMETRI-1][Greece's predecessors in 2000]{\label{fig:gr00}\includegraphics[width=0.24\textwidth, keepaspectratio]{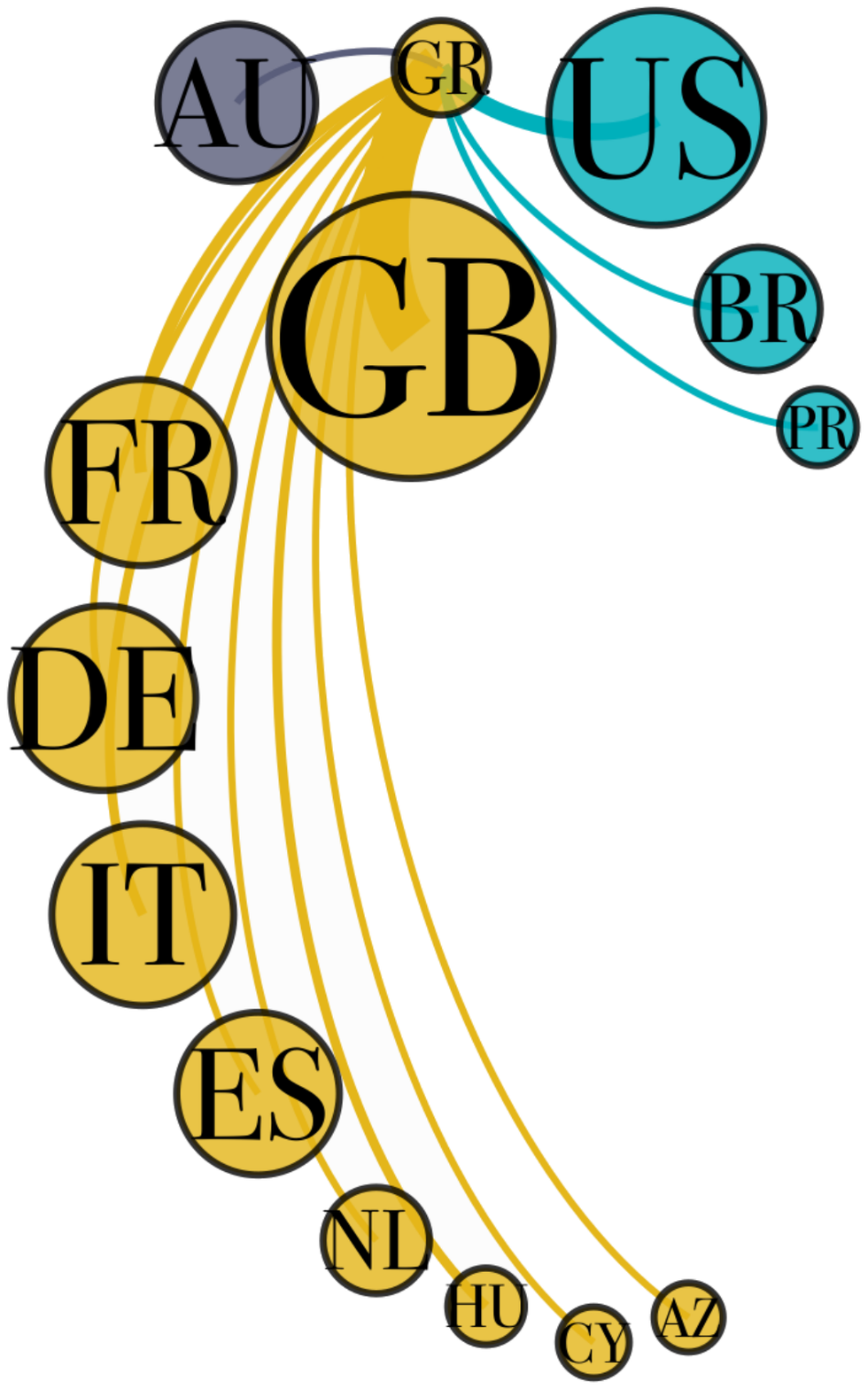}}
\subfigure[PARAMETRI-1][Greece's predecessors in 2016]{\label{fig:gr16}\includegraphics[width=0.24\textwidth, keepaspectratio]{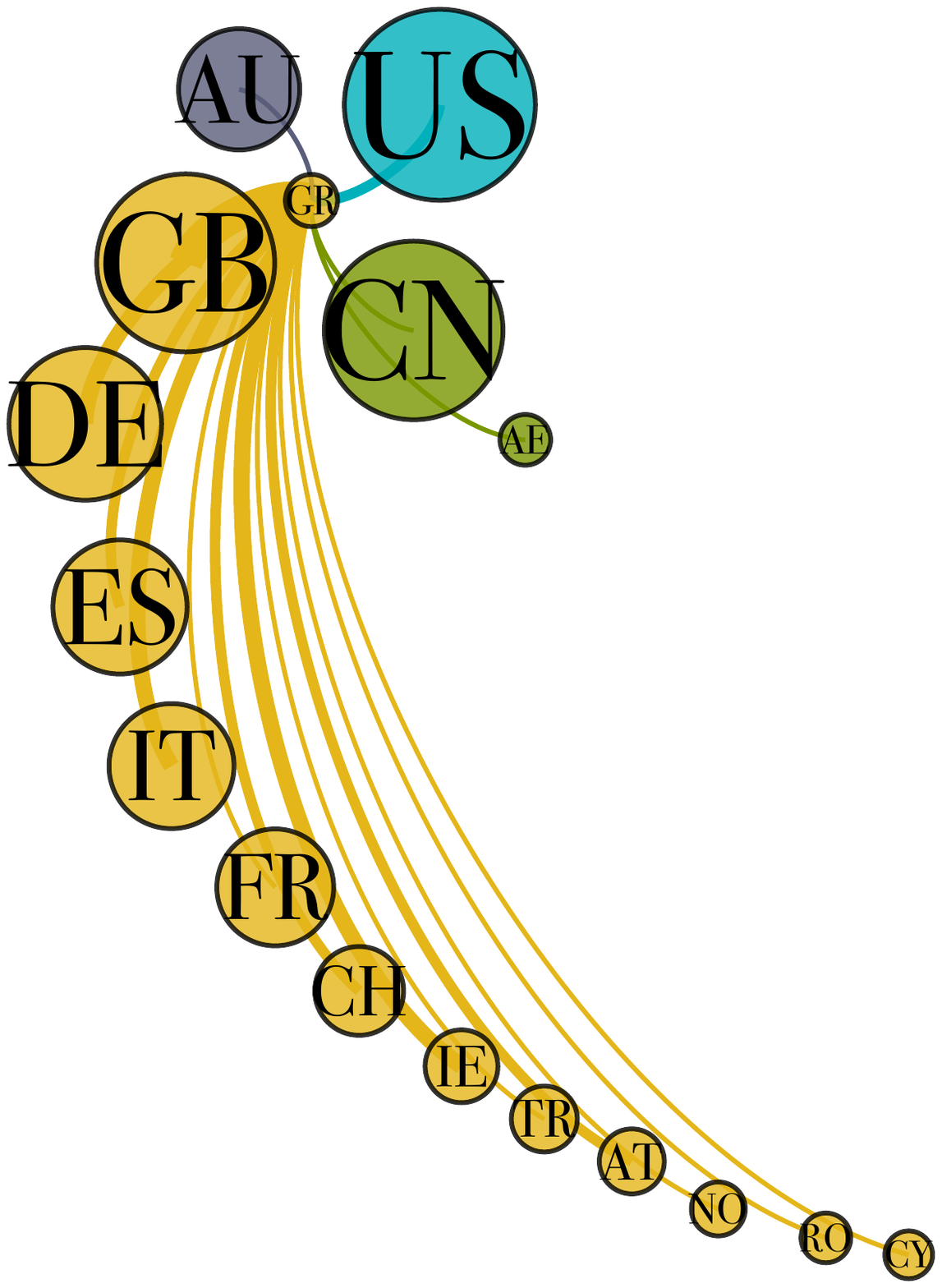}}
\subfigure[PARAMETRI-1][South Korea's successors in 2000]{\label{fig:sk00}\includegraphics[width=0.24\textwidth, keepaspectratio]{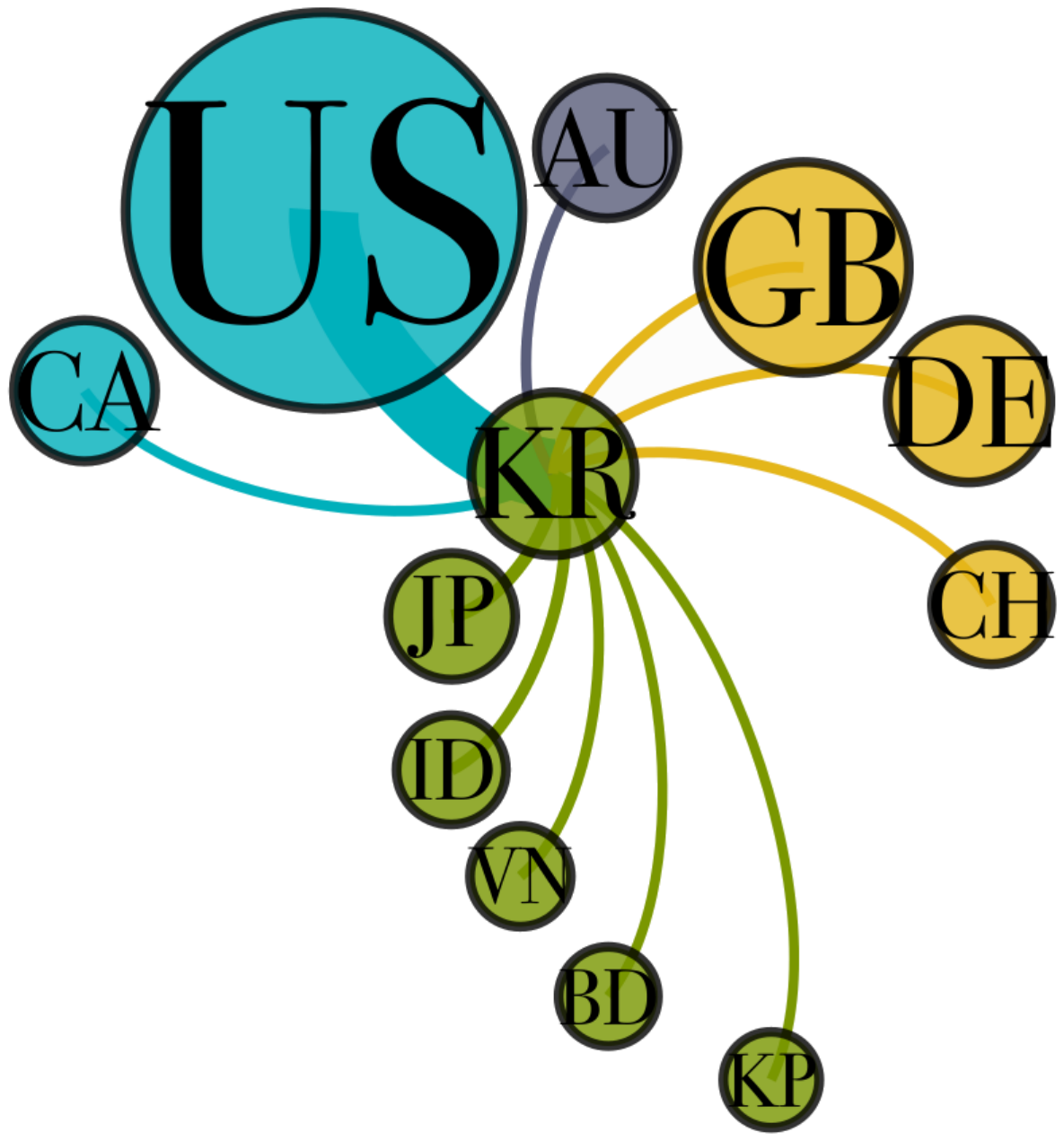}}
\subfigure[PARAMETRI-1][South Korea's successors in 2016]{\label{fig:sk16}\includegraphics[width=0.24\textwidth, keepaspectratio]{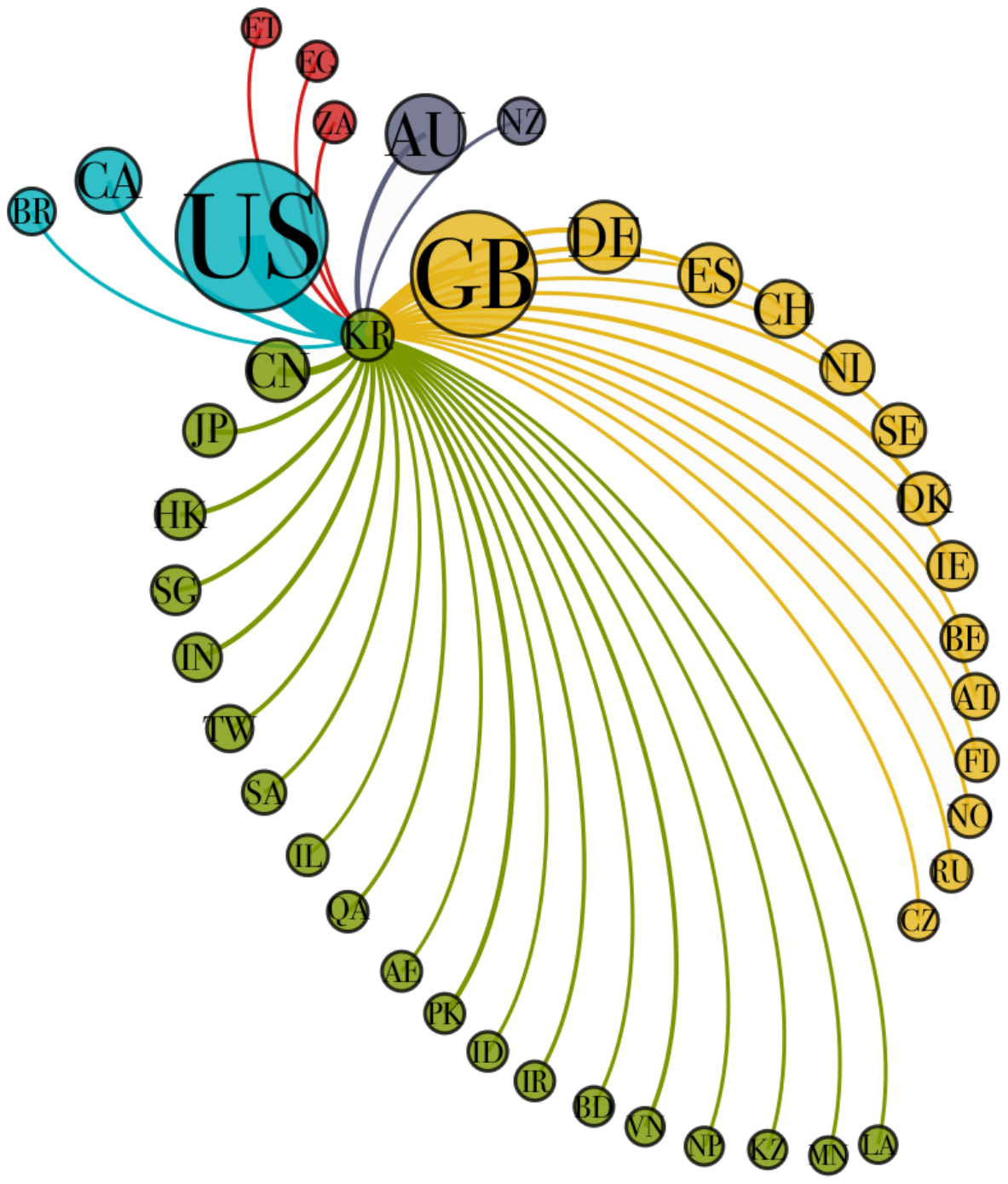}}
\subfigure[PARAMETRI-1][Peru's predecessors in 2000]{\label{fig:pe00}\includegraphics[width=0.24\textwidth, keepaspectratio]{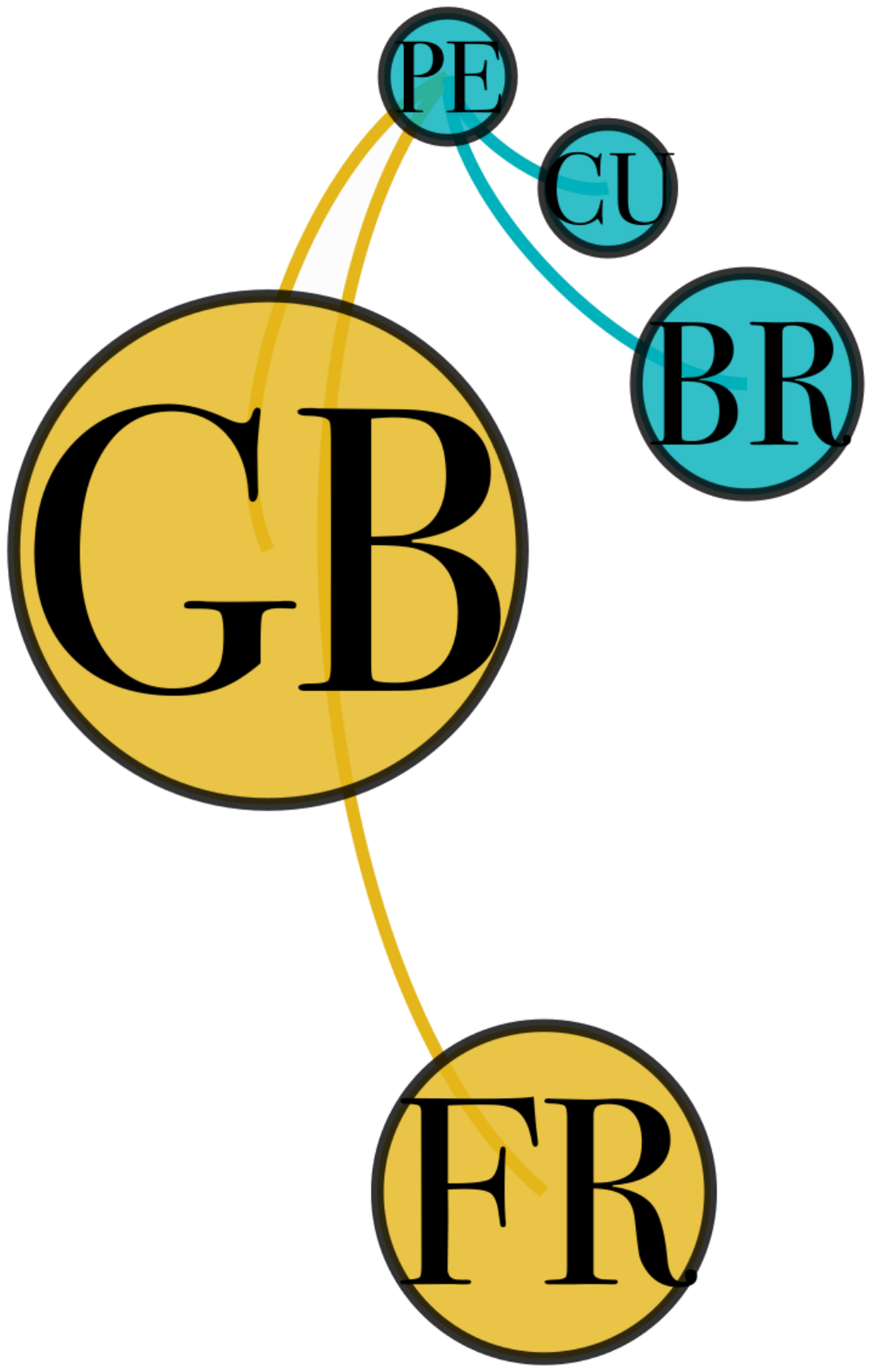}}
\subfigure[PARAMETRI-1][Peru's predecessors in 2016]{\label{fig:pe16}\includegraphics[width=0.24\textwidth, keepaspectratio]{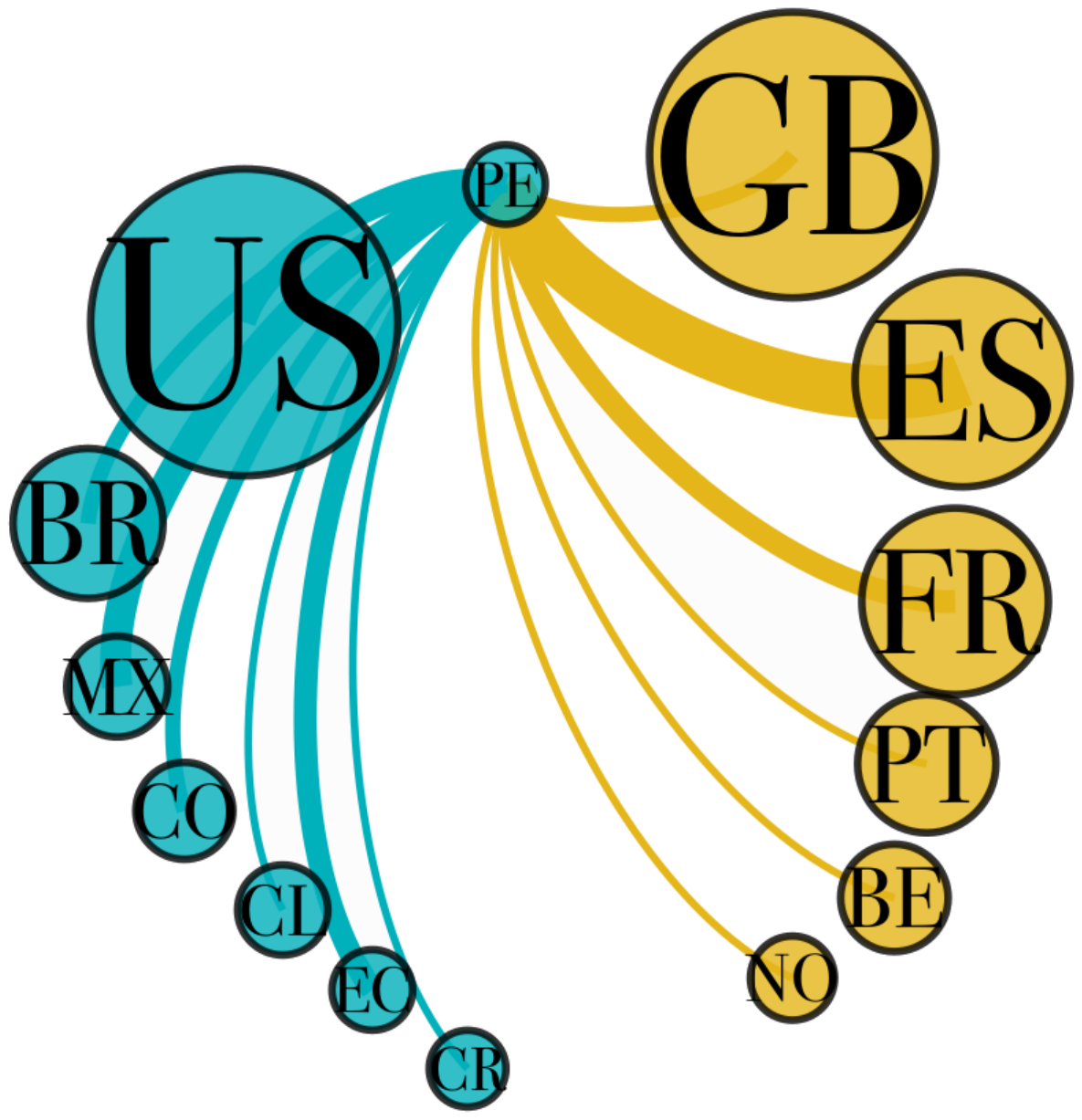}}
\subfigure[PARAMETRI-1][Denmark's successors in 2000]{\label{fig:dk00}\includegraphics[width=0.24\textwidth, keepaspectratio]{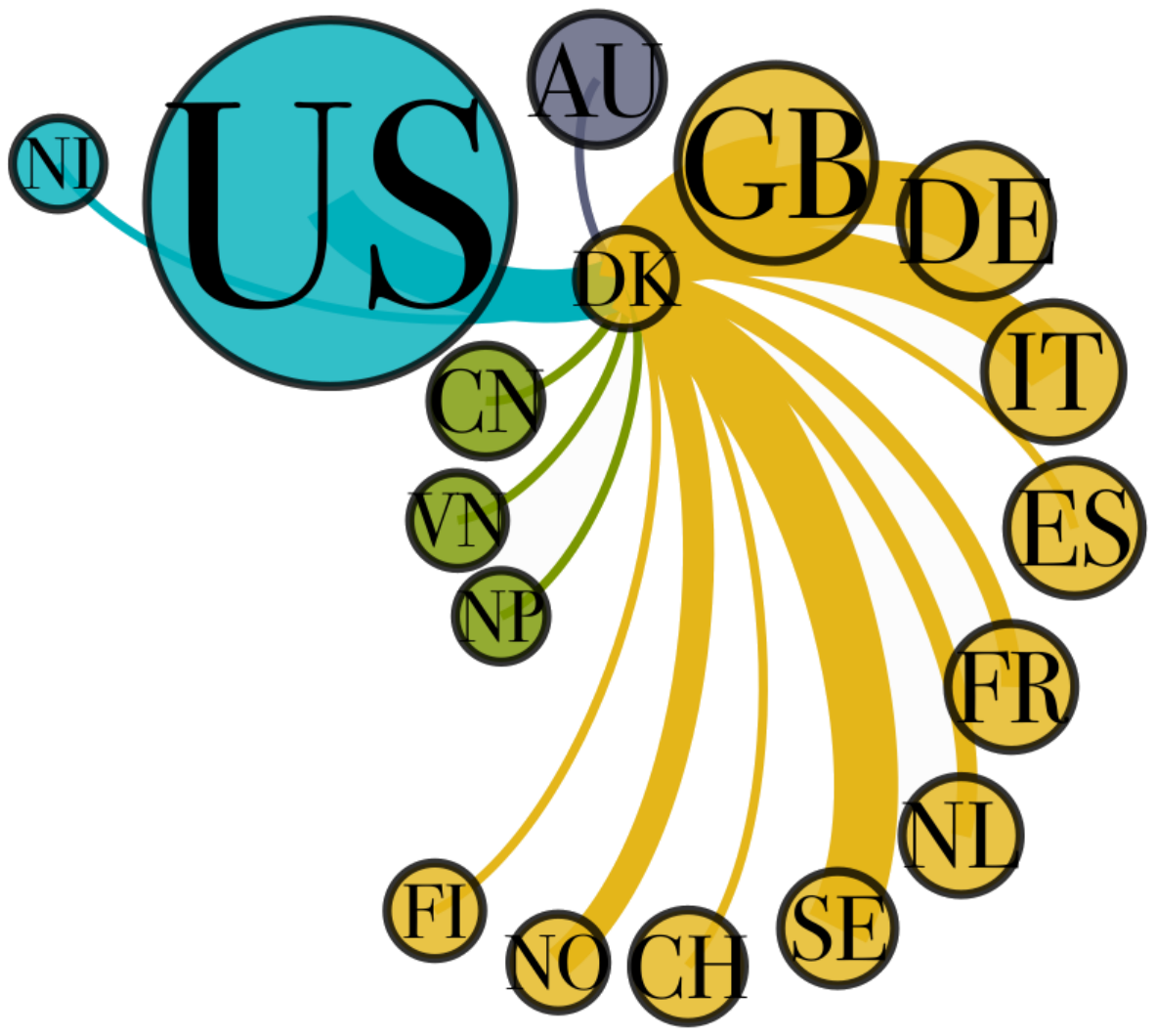}}
\subfigure[PARAMETRI-1][Denmark's successors in 2016]{\label{fig:dk16}\includegraphics[width=0.24\textwidth, keepaspectratio]{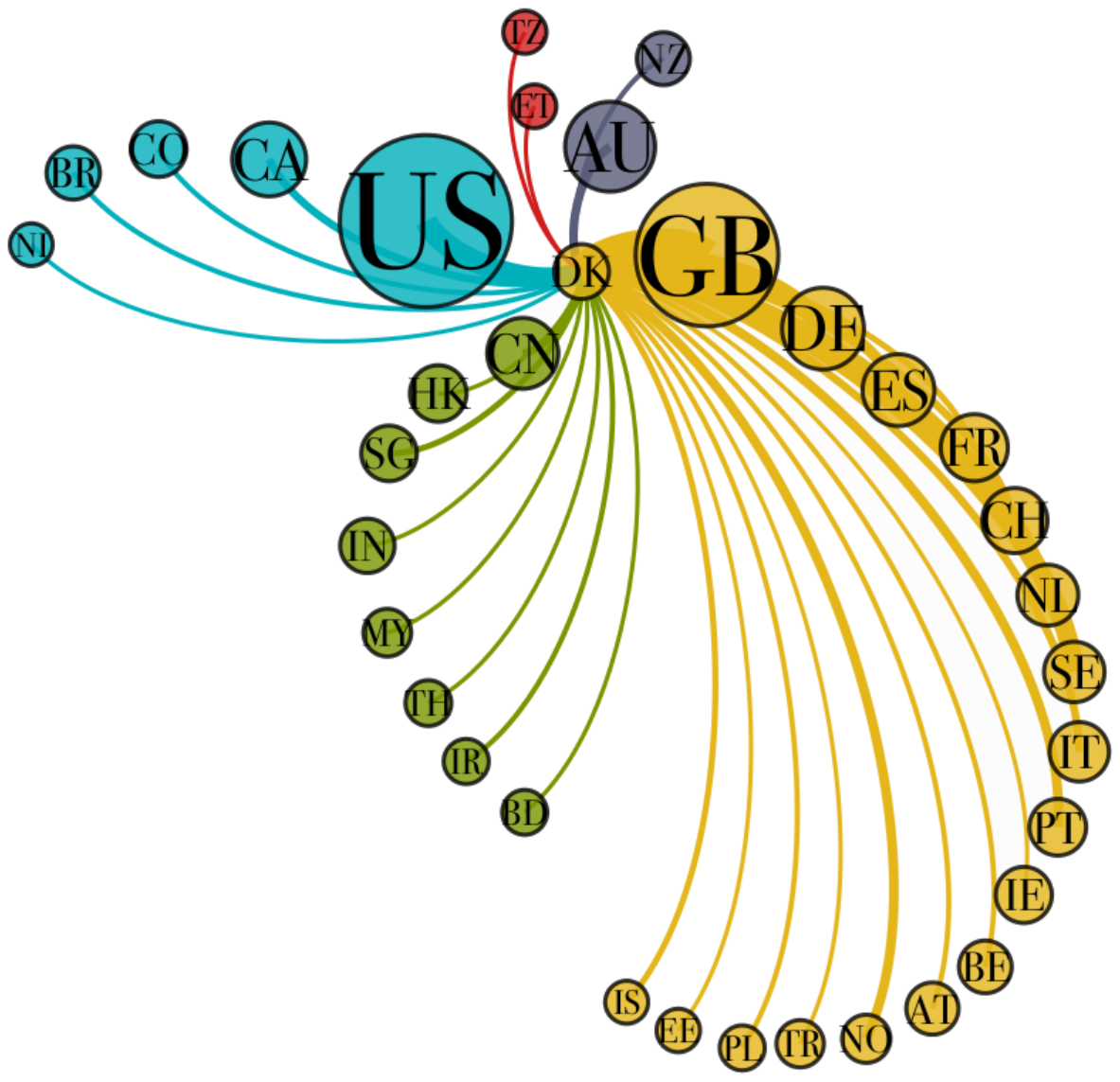}}
\caption{Gini values for weights edges distributions of Greece (GR) and Peru (PE) predecessors (a), and of South Korea (KR) and Denmark (DK) successors (b). We also drew partial ego-networks for the same countries (c-j) in 2000 and 2016: node sizes scale over authority (hub) values for the receiving  (provider) countries; edge thickness is proportional to weights; colours follow continent schema as in Fig.~\ref{fig:hits}.} 
\label{fig:case_studies}
\end{figure}

\subsection{Spotting the impact of heterogeneity with case studies}
To give more concreteness to our discussion, we extrapolate some case studies from the network. Focusing on the nodes that constantly appear in the network backbones created in different years with $\alpha$ equal to 0.2, we plot in Fig.~\ref{fig:ranking_flow} to show the countries that change distinctly their positions in the hub and authority rankings from 2000 to 2016. Among those, we keep also United States and Italy, that are among the countries with less variation in their position, for the sake of comparison.

In the authority rank, for example, Peru and Greece's patterns emerge significantly: Greece loses 21 positions while Peru gains 27. Fig.~\ref{fig:gini_gr_pe} shows how the Gini coefficient of the edges' weight distributions back up both trends: it decreases among the Greece's predecessor edges, and it increases for the Peru's. For example, in 2000 Greece received a lot more of incoming researcher from the United Kingdom (GB) with respect to 2016: see Figures~\ref{fig:gr00} and \ref{fig:gr16}. This could be dependent on a change of 'homecoming' habits: maybe more researchers were able to return to Greece after a period abroad in 2000 rather then 2016. At the same time, it is clear (see Fig.~\ref{fig:gini_gr_pe}) how Peru's predecessors increasingly contributed to incoming streams of uneven strengths over the years, as shown in Fig.~\ref{fig:pe00} and \ref{fig:pe16}. 

W.r.t. the hubs ranking and its variations from 2000 to 2016, we focus on Denmark, that gains an upper position, and to South Korea, that loses some positions (see Fig.~\ref{fig:hub_flow}). Once again, the correlation with the heterogeneity dynamics can be easily spotted in Fig.~\ref{fig:gini_sk_dk}. Focusing on ego-networks again, Denmark's successor edges show an increasing unbalance in their weight distributions (see Figures~\ref{fig:dk00} and \ref{fig:dk16}). On the contrary, South Korea shows a small but significant decreasing heterogeneity among its successors' edges (see Figures~\ref{fig:sk00} and \ref{fig:sk16}). Here the effect is less pronounced than in the other case studies, and at first glance the difference between Fig.~\ref{fig:sk00} and \ref{fig:sk16} can be misleading: in 2016, South Korea exhibits a much wider range of successor countries in its ego network, and this may be incorrectly interpreted as the emergence of a larger heterogeneity. However, we recall that we are referring to edges weights distributions; in fact, in 2000, an out-of-scale outgoing flow to United States is observed, causing an higher Gini coefficient. Conversely, in 2016, a more homogeneous pattern characterises South Korea's ego network, despite a growing number of successor countries. 

Finally, we wish to stress that Fig.~\ref{fig:case_studies} remarks the presence of different heterogeneity's layers: one layer is characterised by different hub and authority scores, and another layer shows not homogeneous in and out strength distributions. If heterogeneity is a signal of complexity, we can observe once again that the interplay between local and global patterns cannot be neglected to identify constantly changing dynamics and to define future scientific mobility strategies.

\section{Conclusions and Future Works}
\label{sec:conclusions}
In this work, we study international migrations of researchers, scientists, and academics using a complex network based approach. This is a data driven study which due to the dataset bias cannot be considered definitive. We mainly focus on proposing a methodology to be applied to data extracted from the \orcid platform to find a measure to quantify the phenomenon of the brain drain.

First of all, we discarded the adoption of simplistic measures that take into account only local measurements of scientists moving in or out, because they lead to rankings that change dramatically from one year to another. As a consequence, we propose to preserve the complexity of the migration ecosystem with adequate measures, that also maintain the dual nature of a country as both an importer and an exported of researchers. Therefore, we model the scientific migration by means of a temporal weighted directed network and employ the \hits algorithm with the intent of catching the interplay between streams of incoming and outgoing researchers from a global prospective.
We also investigate the local characteristics of successors of hubs and predecessors of authorities to dive deeper into the motivations that establish hubs and authorities.

Our findings identify different positions occupied by the main player in the \smn, as shown in Tables~\ref{tab:hits_authority}, Tables~\ref{tab:hits_hub}, and in~\ref{sec:C} for the complete 2016 rankings.
China, United States, and United Kingdom are identified as the leading provider countries during the whole time domain: they never fall below the fifth position.
India and Canada, followed by various of European countries, i.e., Germany, Italy, Spain, and France, consistently position after the three leading countries with few fluctuations during the years.
South Korea and Russia follow instead negative trends.
South Korea is the fifth hub in the \smn during 2000, then loses ten positions by 2016.
About the authority score, United States have the best performance during the whole time horizon, while United Kingdom always classifies \nth{2}.
Germany generally occupies the \nth{3} position in early 2000, before the growth of Australia.
Similarly to the hub score, after the top-4 positions, there is a series of European countries such as Spain, France, and Italy, together with Canada and China. 
Interestingly, among the best receivers, there are Asiatic countries that are not identified as good hubs, e.g., South Korea, Singapore, and Hong Kong, suggesting important efforts in attracting researchers from all over the world and investing for the return of whom left the countries. These dynamics deserve to be further analysed for uncovering latent causes and factors by the inclusion of complementary sources, e.g., local regulations, political alliances, investments in research, development, and education.

At the same time, the evolution of hubs and authorities' scores over time, alongside their relative discrepancy, and other network measures, suggest that local policies can buck the trend, as testified by the Gini coefficient. 

Also, Gini coefficient decreases as $h$ and $a$ decrease, as Figures~\ref{gin_succ_cm} and~\ref{gin_pred_cm} attest.
Complexity in terms of migration patterns seems to co-exist in the best positions of the hub and the authority rankings, in analogy with the economic framework, so that successful countries are extremely diversified in products export~\cite{tacchella2012new}.

Ranking by means of hubs and authorities scores it is insightful, but just a preliminary step toward a more refined analysis. As future work, we plan to expand the study carried out so far by tackling the correlation between hub and authority scores with respect to metrics of research/academic success and economic indicators, as in~\cite{fernandez2016productivity}; even though not very high due to the presence of countries of high \textsf{GDP} showing poor performances in terms of hub or authority ranking, as also discussed in~\cite{van2012global}, where the relationship between science and investments shows complex behaviours.
Furthermore, we plan to restrict the analysis to a specific geographical region (e.g., Europe) to study migrations at smaller granularity (e.g., cities) or, according to specific science fields in order to understand where skills actually move, and by different career stages, education or employment.

Finally, we plan to adapt our methodology to evolving datasets that grow over time, to deliver a more precise picture as the information increases. In particular, we would like to design a permanent observatory of the \smn, that can keep track over the year of the multiple aspects of this rich and complex phenomenon and this work is an important step in that direction. Moreover, it is important to mention that the ambition of our proposed methodology is that hub and authority scores will be considered in forthcoming biblio-metric observatories and studies, to exploit the interplay of incoming and outgoing scientific migration flows, to better understand the role of single countries in a world-wide interconnected ecosystem.

It would be of interest to replicate our analysis on other data sources to confirm/integrate our results, and to keep updating the analysis to ever evolving global and local scenarios.

\section*{List of abbreviations}
\label{sec:abbreviations}
\begin{abbrv}
\item[\textsf{SMN}]			scientific migration network
\item[\hits]			hyperlink-induced topic search
\item[\textsf{GDP}]			gross domestic product
\end{abbrv}

\section*{Declarations}

\subsection*{Funding}
AU acknowledges support from the Lagrange Project of the Institute for Scientific Interchange Foundation (ISI Foundation), funded by Fondazione Cassa di Risparmio di Torino (Fondazione CRT).
Additionally, authors have been partially supported by the project ``Analisi di Reti Complesse e di Sistemi Socio-Tecnologici'', funded by University of Turin.
\subsection*{Conflict of interest/Competing interests}
The authors declare that they have no competing interests.
\subsection*{Availability of data and materials}
The ``ORCID migrations by person'' dataset analyzed during the current study is available in the Dryad Digital Repository, \url{https://doi.org/10.5061/dryad.48s16}~\cite{dryad_48s16}. Also, networked data used for our analysis, will be made available on public repository upon paper's acceptance.
\subsection*{Code availability} Code will be made available on public repository upon paper's acceptance.
\subsection*{Author's contributions}
\emph{Conceptualization:} G. Ruffo; 
\emph{Data curation:} A. Urbinati, E. Galimberti;
\emph{Formal analysis:} E. Galimberti, A. Urbinati; 
\emph{Methodology:} A. Urbinati, E. Galimberti, G. Ruffo;
\emph{Validation:} G. Ruffo, A. Urbinati; 
\emph{Visualization:} A. Urbinati, E. Galimberti;
\emph{Writing - original draft: } A. Urbinati, E. Galimberti;
\emph{Writing - review \& editing:} A. Urbinati, G. Ruffo.

\section*{Acknowledgements}
The authors would like to acknowledge Alessandro Flammini and Roberta Sinatra for feedbacks and insightful conversations on a preliminary versions of this work.

\bibliography{mybibfile}

\begin{appendix}

\section{Network Model}
\label{sec:A}
In Table~\ref{tab:summary}, we show some basic network statistics, grouped by year. For each year $y \in [2000, \ldots 2016]$ we show the number of \textit{nodes}, i.e., countries that occur as a source or as a destination in that year at least once ($|V_y|$), the number of \textit{links} established during that year between countries ($|E_y|$), and the related following measures: the \textit{density} of the network ($d = \frac{2|E_y|}{|V_y|(|V_y| - 1)}$); the \textit{reciprocity}, i.e., the ratio of the number of edges pointing in both directions to the total number of edges in the graph ($r=\frac{|e =
 (i,j): (j, i))\in E_y|}{|E_y|}$); the size of the Strongly Connected Component (\textit{SCC}); and the diameter of the network, i.e., the length of the longest path among the shortest ones. 

\begin{table}
\centering
\caption{Summary of some basic network statistics, grouped by year.}
\label{tab:summary}
\begin{tabularx}{0.95\columnwidth}{cXXXXXc}
\toprule
year &  \# nodes &  \# links &  density &  reciprocity &  SCC &  diameter \\
\midrule
\textbf{2000}&170&1341&0.047&0.552&124&5\\
\textbf{2001}&165&1396&0.052&0.549&123&6\\
\textbf{2002}&170&1476&0.051&0.562&127&5\\
\textbf{2003}&168&1530&0.055&0.571&127&6\\
\textbf{2004}&174&1661&0.055&0.580&136&4\\
\textbf{2005}&172&1815&0.062&0.608&140&5\\
\textbf{2006}&180&1942&0.060&0.582&148&5\\
\textbf{2007}&187&2103&0.060&0.602&144&5\\
\textbf{2008}&190&2259&0.063&0.602&146&5\\
\textbf{2009}&190&2457&0.068&0.597&156&5\\
\textbf{2010}&190&2514&0.070&0.621&152&4\\
\textbf{2011}&198&2655&0.068&0.627&164&5\\
\textbf{2012}&198&2916&0.075&0.634&167&4\\
\textbf{2013}&203&3041&0.074&0.622&172&4\\
\textbf{2014}&206&3035&0.072&0.611&171&5\\
\textbf{2015}&197&2872&0.074&0.604&163&4\\
\textbf{2016}&173&2133&0.072&0.625&135&4\\
\bottomrule
\end{tabularx}
\end{table}

\begin{figure}[ht!]
\centering
\includegraphics[width=0.95\columnwidth]{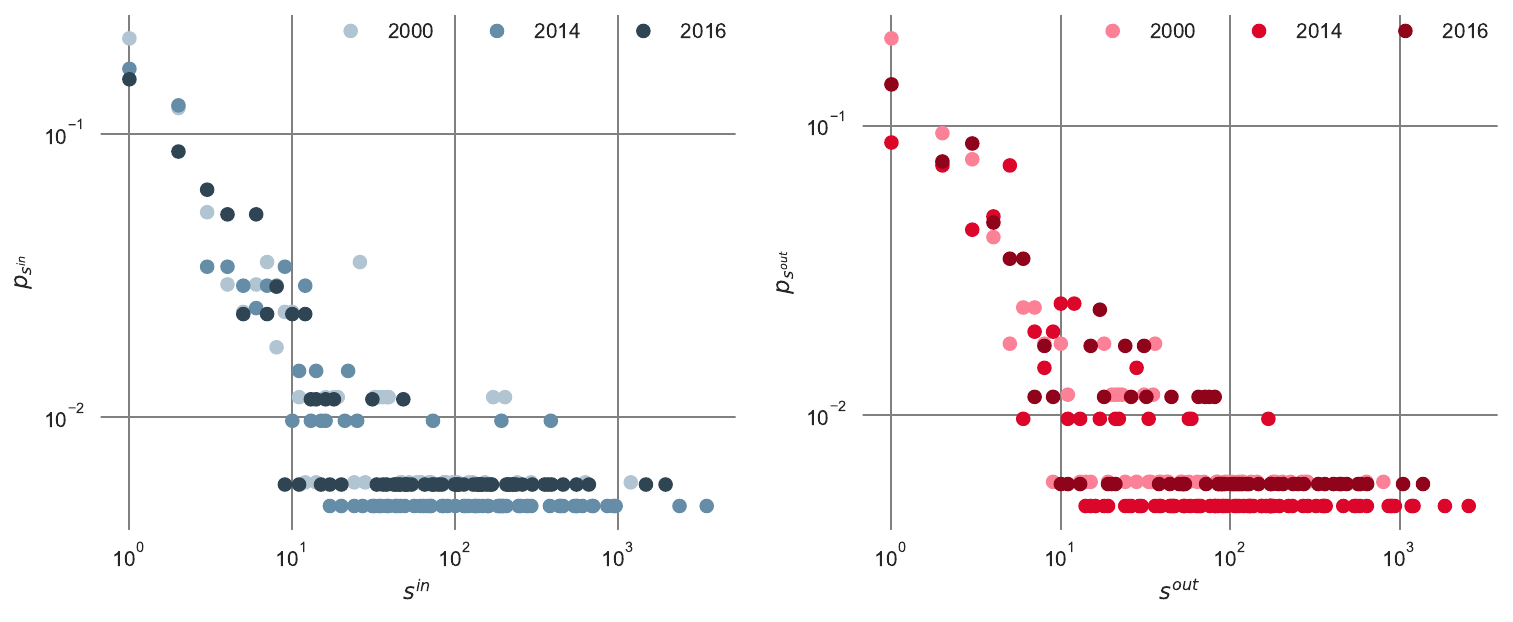}
\caption{In-strength (left) and the out-strength (right) distributions in the \smn in 2000, 2014, and 2016.}
\label{fig:strength_distr}
\end{figure}

We show in Figure~\ref{fig:strength_distr} the  in-strength and the out-strength distributions in the \smn in 2000, 2014, and 2016. Other years are not reported here, but they show a comparable behaviour: the shapes of the distributions are very similar to each other.
Also, there are not notable differences between in-strength and out-strength.
Such distributions will come in handy in the following, as input of configuration models that create random graphs preserving in-strength and out-strength sequences.

\section{Correlations among measures}
\label{sec:B}
\begin{figure}[ht!]
\centering
\includegraphics[width=0.95\columnwidth]{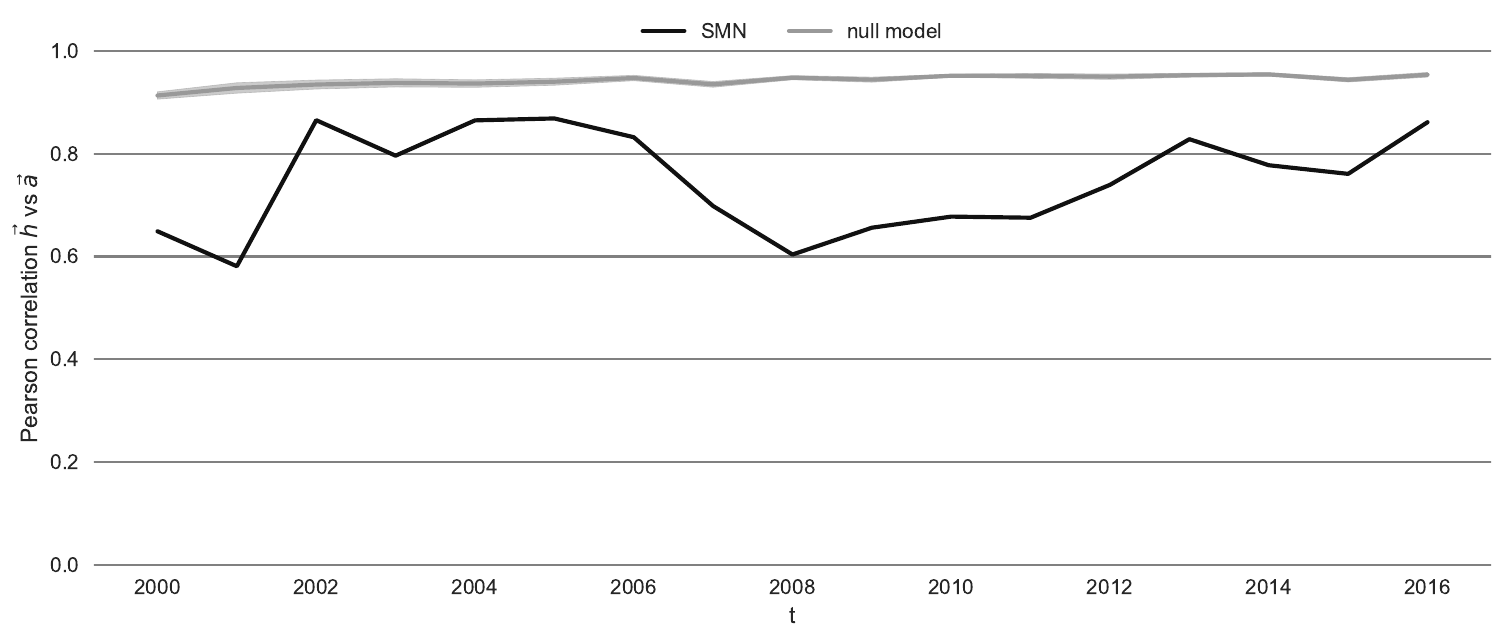}
\caption{Person correlation between $\vec{h}$ and $\vec{a}$ of the \smn and of the null model, for which we report mean and 95\% confidence interval.
$p$-values are smaller than $1.5e\!-\!05$ in all cases.}
\label{fig:hits_corr}
\end{figure}

The correlation between $\vec{h}$ and $\vec{a}$ and the evolution of such correlation is an interesting aspect to take into account.
We show, in Figure~\ref{fig:hits_corr}, the Pearson correlation between $\vec{h}$ and $\vec{a}$ as a function of the year, and compare it to a null model.

The correlation in the original network is strong during the whole time domain, constantly greater than $0.85$.
The null model has even stronger correlation in all years, with small variation between the different configurations.
This means that we should expect more countries of high (low) hub score having also high (low) authority score, and vice versa, in the \smn.
The observed behaviour should then rely on different factors, e.g., local patterns than the strength distribution.
\begin{figure}[ht!]
\centering
\includegraphics[width=0.98\columnwidth]{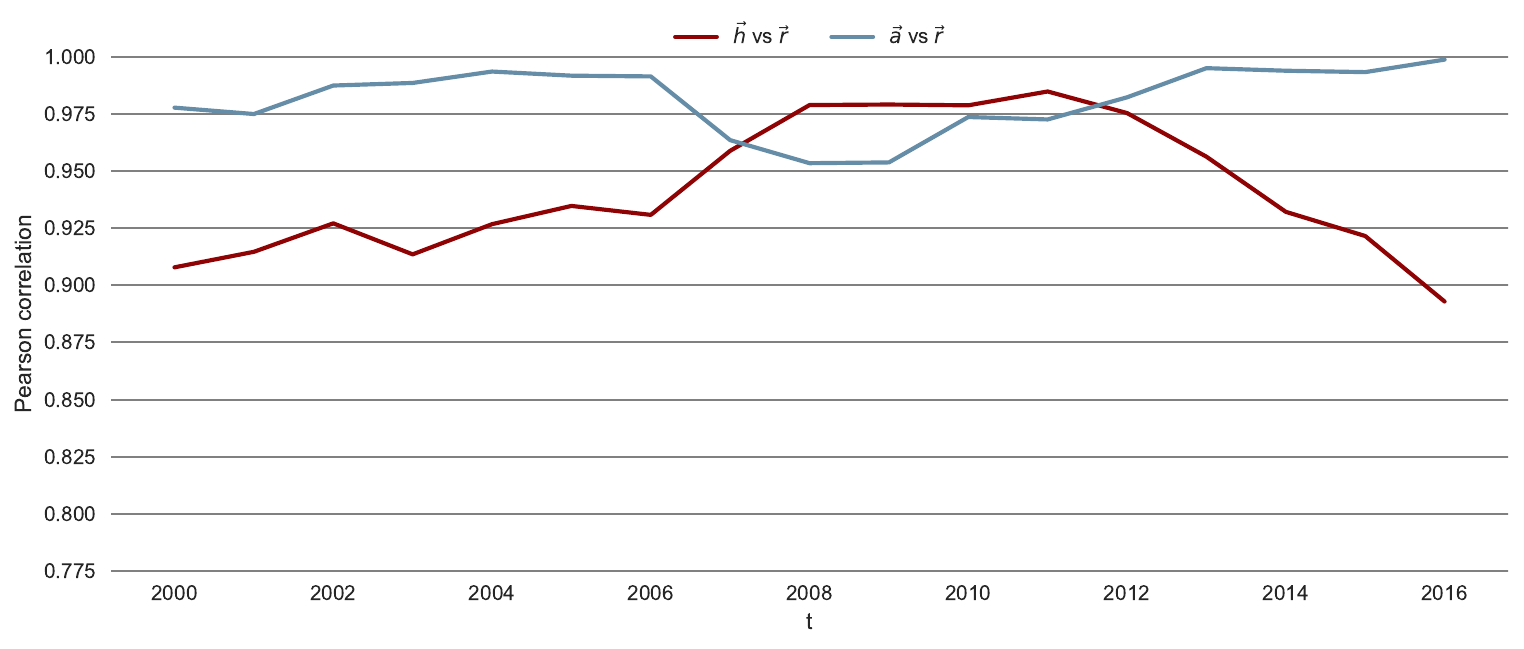}
\caption{Person correlation between $\vec{h}$ and $\vec{a}$, and $\vec{r}$ of the \smn.}
\label{fig:hits_pagerank_corr}
\end{figure}

In order to compare the \hits and the PageRank results, in Figure~\ref{fig:hits_pagerank_corr} we also visualise the Pearson correlation between $\vec{h}$ and $\vec{a}$, and $\vec{r}$.
Interestingly, both $\vec{h}$ and $\vec{a}$ are highly correlated to $\vec{r}$.
$\vec{a}$, in particular, has correlation greater than $0.95$ in all years.
This validates the results obtained by the \hits algorithm that has the advantage of depicting two different aspects of the world countries, providing then more accurate indications.

\section{HITS complete ranking}
2016 rankings of countries according authority and hub scores are reported in this section for illustrative purposes. We are aware that this information will be obsolescent at the time of publication; however this is based on the dataset provided in~\cite{dryad_48s16}, that has been collected from ORCID in 2017 and made available to the community. We claim that temporal scientific migration networks can be built from actual ORCID data and that always up-to-date rankings and accessory information can be explored by the interested user via a Web based dashboard. However, the implementation of such software architecture is beyond the scope of this paper.
\label{sec:C}
\begin{table}[ht]
\renewcommand{\arraystretch}{0.67}
\caption{Ranking of the countries by authority score in 2016.
\label{tab:aut2016}}
\begin{tabular}{@{\hspace{-2cm}}rlrlrlrl}
\toprule

    1 &         United States &      46 &          Iran &      91 &                   Serbia &     136 &                     New Caledonia \\
    2 &        United Kingdom &      47 &        Greece &      92 &                  Albania &     137 &                         Lithuania \\
    3 &             Australia &      48 &        Poland &      93 &  Palestinian Territories &     138 &                            Angola \\
    4 &               Germany &      49 &    Bangladesh &      94 &                  Iceland &     139 &                       Timor-Leste \\
    5 &                Canada &      50 &          Peru &      95 &                  Algeria &     140 &                          Mongolia \\
    6 &                 Spain &      51 &    Luxembourg &      96 &                Mauritius &     141 &                           Bolivia \\
    7 &                 China &      52 &     Sri Lanka &      97 &                 Zimbabwe &     142 &                   Myanmar [Burma] \\
    8 &                France &      53 &          Iraq &      98 &                 Slovakia &     143 &                  Congo [Republic] \\
    9 &           Switzerland &      54 &       Hungary &      99 &                    Malta &     144 &                       Congo [DRC] \\
   10 &           Netherlands &      55 &         Kenya &     100 &                  Tunisia &     145 &                       Afghanistan \\
   11 &                Sweden &      56 &       Nigeria &     101 &               Madagascar &     146 &                      Turkmenistan \\
   12 &                 Japan &      57 &      Ethiopia &     102 &         Papua New Guinea &     147 &                           Curacao \\
   13 &                 Italy &      58 &       Vietnam &     103 &                Nicaragua &     148 &                  French Polynesia \\
   14 &               Denmark &      59 &         Nepal &     104 &      Trinidad and Tobago &     149 &                            Belize \\
   15 &              Portugal &      60 &    Kazakhstan &     105 &                 Paraguay &     150 &                             Libya \\
   16 &             Hong Kong &      61 &       Uruguay &     106 &                   Zambia &     151 &                        Uzbekistan \\
   17 &               Ireland &      62 &   Philippines &     107 &               Mozambique &     152 &                      Cote dIvoire \\
   18 &              Colombia &      63 &       Lebanon &     108 &                     Laos &     153 &                        Montenegro \\
   19 &             Singapore &      64 &         Sudan &     109 &                   Bhutan &     154 &                              Togo \\
   20 &                 India &      65 &        Uganda &     110 &                   Rwanda &     155 &                             Tonga \\
   21 &           South Korea &      66 &       Estonia &     111 &               Azerbaijan &     156 &  Saint Vincent \\
   22 &                Brazil &      67 &    Costa Rica &     112 &                 Cameroon &     157 &                           Burundi \\
   23 &           New Zealand &      68 &       Romania &     113 &                  Senegal &     158 &                        Guadeloupe \\
   24 &               Belgium &      69 &         Ghana &     114 &                   Brunei &     159 &                             Niger \\
   25 &                Taiwan &      70 &        Cyprus &     115 &                  Grenada &     160 &                         Swaziland \\
   26 &               Austria &      71 &     Guatemala &     116 &                  Jamaica &     161 &                        Kyrgyzstan \\
   27 &                Mexico &      72 &      Slovenia &     117 &                    Syria &     162 &                            Guyana \\
   28 &          Saudi Arabia &      73 &      Honduras &     118 &               Cape Verde &     163 &             Saint Kitts and Nevis \\
   29 &                 Chile &      74 &        Panama &     119 &                  Morocco &     164 &                           Belarus \\
   30 &               Finland &      75 &      Tanzania &     120 &      Antigua and Barbuda &     165 &                         Greenland \\
   31 &                Norway &      76 &         Benin &     121 &                  Bahrain &     166 &                           Bahamas \\
   32 &              Malaysia &      77 &     Venezuela &     122 &             Burkina Faso &     167 &            British Virgin Islands \\
   33 &               Ecuador &      78 &       Ukraine &     123 &       Dominican Republic &     168 &                             Yemen \\
   34 &                Turkey &      79 &       Croatia &     124 &                  Somalia &     169 &                          Maldives \\
   35 &          South Africa &      80 &   Puerto Rico &     125 &            Faroe Islands &     170 &                            Guinea \\
   36 &                Israel &      81 &        Kuwait &     126 &                   Gambia &     171 &                           Eritrea \\
   37 &                Russia &      82 &          Oman &     127 &                  Armenia &     172 &                           Liberia \\
   38 &                 Qatar &      83 &      Bulgaria &     128 &                     Cuba &     173 &                 Macedonia [FYROM] \\
   39 &              Thailand &      84 &         Macau &     129 &                 Cambodia &      &                                  \\
   40 &                 Egypt &      85 &        Jordan &     130 &                    Gabon &      &                                  \\
   41 &  United Arab Emirates &      86 &   El Salvador &     131 &              Isle of Man &      &                                  \\
   42 &              Pakistan &      87 &      Botswana &     132 &              South Sudan &      &                                  \\
   43 &        Czech Republic &      88 &  Sierra Leone &     133 &                 Guernsey &      &                                  \\
   44 &             Indonesia &      89 &          Fiji &     134 &                   Latvia &      &                                  \\
   45 &             Argentina &      90 &        Malawi &     135 &                     Chad &      &                                  \\
\bottomrule
\end{tabular}
\end{table}
\begin{table}[ht]
\renewcommand{\arraystretch}{0.67}
\caption{Ranking of the countries by hub score in 2016.
\label{tab:hub2016}}
\begin{tabular}{@{\hspace{-2cm}}rlrlrlrl}
\toprule
    1 &   United States &      46 &                 Qatar &      91 &                            Kuwait &     136 &               Guinea \\
    2 &           China &      47 &             Argentina &      92 &                      Cote dIvoire &     137 &    Macedonia [FYROM] \\
    3 &  United Kingdom &      48 &               Vietnam &      93 &                            Panama &     138 &           Azerbaijan \\
    4 &         Germany &      49 &              Thailand &      94 &                          Cameroon &     139 &              Eritrea \\
    5 &           India &      50 &               Hungary &      95 &                            Brunei &     140 &             Cambodia \\
    6 &           Spain &      51 &           Puerto Rico &      96 &                         Guatemala &     141 &              Senegal \\
    7 &          Canada &      52 &               Ecuador &      97 &           Palestinian Territories &     142 &           Guadeloupe \\
    8 &           Italy &      53 &             Sri Lanka &      98 &                            Rwanda &     143 &                Syria \\
    9 &       Australia &      54 &                 Ghana &      99 &                           Armenia &     144 &            Greenland \\
   10 &          France &      55 &  United Arab Emirates &     100 &                            Uganda &     145 &               Angola \\
   11 &     Netherlands &      56 &           Philippines &     101 &                            Guyana &     146 &          Timor-Leste \\
   12 &          Brazil &      57 &                  Peru &     102 &                      Sierra Leone &     147 &        Faroe Islands \\
   13 &     Switzerland &      58 &                 Kenya &     103 &                       South Sudan &     148 &           Madagascar \\
   14 &        Portugal &      59 &                 Nepal &     104 &                           Liberia &     149 &                 Oman \\
   15 &     South Korea &      60 &               Lebanon &     105 &                        Kyrgyzstan &     150 &               Zambia \\
   16 &          Sweden &      61 &             Venezuela &     106 &                         Swaziland &     151 &                Benin \\
   17 &           Japan &      62 &               Romania &     107 &                        Cape Verde &     152 &              Burundi \\
   18 &         Denmark &      63 &               Ukraine &     108 &             Saint Kitts and Nevis &     153 &                Niger \\
   19 &         Ireland &      64 &                Jordan &     109 &            British Virgin Islands &     154 &         Turkmenistan \\
   20 &         Belgium &      65 &            Costa Rica &     110 &  Saint Vincent &     155 &                Gabon \\
   21 &          Turkey &      66 &                Serbia &     111 &                             Malta &     156 &              Jamaica \\
   22 &       Singapore &      67 &              Ethiopia &     112 &                          Slovakia &     157 &               Belize \\
   23 &         Austria &      68 &                  Cuba &     113 &                           Belarus &     158 &         Burkina Faso \\
   24 &            Iran &      69 &            Luxembourg &     114 &                          Maldives &     159 &               Gambia \\
   25 &          Greece &      70 &               Morocco &     115 &                             Libya &     160 &     Papua New Guinea \\
   26 &          Mexico &      71 &            Kazakhstan &     116 &                              Fiji &     161 &        New Caledonia \\
   27 &         Finland &      72 &               Estonia &     117 &                           Somalia &     162 &                Tonga \\
   28 &       Hong Kong &      73 &                 Sudan &     118 &                          Botswana &     163 &           Montenegro \\
   29 &           Egypt &      74 &    Dominican Republic &     119 &                  Congo [Republic] &     164 &          Afghanistan \\
   30 &        Colombia &      75 &              Tanzania &     120 &                  French Polynesia &     165 &  Antigua and Barbuda \\
   31 &    Saudi Arabia &      76 &               Tunisia &     121 &                       El Salvador &     166 &           Uzbekistan \\
   32 &          Taiwan &      77 &              Bulgaria &     122 &                           Bolivia &     167 &             Mongolia \\
   33 &    South Africa &      78 &               Croatia &     123 &                           Bahrain &     168 &                 Chad \\
   34 &          Israel &      79 &                Latvia &     124 &                              Laos &     169 &                 Togo \\
   35 &     New Zealand &      80 &              Zimbabwe &     125 &                        Mozambique &     170 &            Mauritius \\
   36 &      Bangladesh &      81 &               Grenada &     126 &                            Bhutan &     171 &              Curacao \\
   37 &        Malaysia &      82 &             Nicaragua &     127 &                          Honduras &     172 &             Guernsey \\
   38 &          Russia &      83 &                  Iraq &     128 &                       Congo [DRC] &     173 &          Isle of Man \\
   39 &           Chile &      84 &                Malawi &     129 &                           Algeria &      &                     \\
   40 &        Pakistan &      85 &                Cyprus &     130 &               Trinidad and Tobago &      &                     \\
   41 &         Nigeria &      86 &               Uruguay &     131 &                           Iceland &      &                     \\
   42 &          Poland &      87 &              Slovenia &     132 &                           Albania &      &                     \\
   43 &  Czech Republic &      88 &             Lithuania &     133 &                             Yemen &      &                     \\
   44 &       Indonesia &      89 &               Bahamas &     134 &                             Macau &      &                     \\
   45 &          Norway &      90 &       Myanmar [Burma] &     135 &                          Paraguay &      &                     \\
\bottomrule
\end{tabular}
\end{table}
\end{appendix}

\end{document}